\documentclass{article}
\usepackage[T1]{fontenc}
\usepackage{amssymb}
\usepackage{amsfonts}
\usepackage{amsmath}
\usepackage{epsfig}
\usepackage{graphicx}
\title{Estimate of the free energy difference in mechanical systems from work
        fluctuations: experiments and models}
\author{F. Douarche, S. Ciliberto, A. Petrosyan\\
        Laboratoire de Physique de l'ENS Lyon -- CNRS UMR 5672\\
        46, All\'ee d'Italie -- 69364 Lyon Cedex 07, France}

\def\ga{\gamma}
\def\om{\omega}
\def\lam{\lambda}

\def\d{\,\mathrm{d}}
\begin{document}
\maketitle
\begin{abstract}
The work fluctuations of an oscillator in contact with a heat
reservoir and driven out of equilibrium by an external force are
studied experimentally. The oscillator dynamics  is modeled by a
Langevin equation. We find both experimentally and theoretically
that, if the driving force does not change the equilibrium
properties of the thermal fluctuations of this mechanical system,
the free energy difference $\Delta F$ between two equilibrium
states can be exactly computed using the Jarzynski equality (JE)
and the Crooks relation (CR) \cite{jarzynski1, crooks1,
jarzynski2}, independently of the time scale and amplitude of the
driving force. The applicability limits for the JE and CR at very
large driving forces are discussed. Finally, when the work
fluctuations are Gaussian, we propose an alternative  method to
compute $\Delta F$ which can be safely applied, even in cases
where the JE and CR might not hold. The results of this paper are
useful to compute $\Delta F$ in complex systems such as the
biological ones.
\end{abstract}
\section{Introduction}
A precise characterization of the dynamics of mesoscopic and
macroscopic systems is a very important problem for applications
in nanotechnologies and biophysics. While fluctuations always play
a negligible role in large systems, their influence may become
extremely important in small systems driven out of equilibrium.
Such is the case for nanoengines and biological processes, where
the characteristic amount of energy transferred by fluctuations
can be of the same order as that which operates the device. In
these small out of equilibrium systems, dominated by thermal
fluctuations, a precise estimation of the free energy difference
$\Delta F$ between two equilibrium states $A$ and $B$ is extremely
useful to increase our knowledge of the underlying physical
processes which control their dynamical behaviour. It is well
known that $\Delta F$ can be estimated by perturbing a system with
an external parameter $\lam$ and by measuring the work $W$ done to
drive the system from $A$ to $B$. When thermal fluctuations cannot
be neglected $W$ is a strongly fluctuating quantity and $\Delta F$
may in principle be computed using the Jarzynski equality (JE)
\cite{jarzynski1} and the Crooks relation (CR) \cite{crooks1} (see
section 2 for their definition). Indeed the JE and CR take
advantage of these work fluctuations and relate the $\Delta F$ to
the probability distribution function (pdf) of the work performed
on the system to drive it from $A$ to $B$ along any path $\ga$  (
either reversible or irreversible) in the system parameter space.
Numerous derivations of the JE and CR have been produced
\cite{crooks2, jarzynski3, mazonka_jarzynski, crooks3, wojcik} and
JE and CR have been used in order to estimate the free energy of a
stretched DNA molecule \cite{bustamante,ritort}. However the
derivation of JE in the case of strongly irreversible systems has
been criticized in ref.\cite{cohen}. For this reason in a recent
letter \cite{Europhysics}, we have experimentally checked JE and
CR on a very simple and controlled device in order to safely use
these two relations in more complex cases as the biological and
chemical ones, where it is much more difficult (almost impossible)
to verify the results with other methods. Our system is a
mechanical oscillator, in contact with a heat bath, which is
driven out of equilibrium, between two equilibrium states $A$ and
$B$, by a small external force. In ref.\cite{Europhysics} we have
shown that the JE and CR are experimentally accessible and valid
for all the values of the different control parameters. However
one may wonder whether the driving force's switching rate and its
amplitude were fast and large enough.  Therefore  in this paper we
extend the measurements of ref.\cite{Europhysics} till the limits
of the experimental set up.  We find  that even in these extreme
cases the $\Delta F$ is correctly estimated by JE and CR. We also
observe that these experimental results can be exactly derived
from a Langevin equation if one takes into account that,
experimentally,  the properties of the thermal noise are not
changed by the external driving force. These conditions are close
to those pointed out in Ref.\,\cite{cohen} for the validity of the
JE, thus they do not fully alight the theoretical debate. However
the fact of having driven the system at large driving amplitudes
and the study of the corresponding Langevin dynamics allow us to
clearly understand the experimental limits of applicability of the
JE and CR. Independently on the theoretical problems that the
derivation of JE may rise, these limits are of statistical nature
and they strongly reduce the possibility of using  JE and CR
 when the driving energy is much larger than
the thermal energy. As a consequence these intrinsic statistical
limits make the experimental test of the criticisms arisen in
Ref.\,\cite{cohen} very difficult almost impossible. However, when
the fluctuations of $W$ are Gaussian we propose here an
alternative  method to compute $\Delta F$,  which can be useful
even in cases where the JE and the CR could not hold.

The paper is organized as follows. In section 2 we recall the JE
and CR and we discuss the Gaussian case. In section 3 we describe
the experimental setup and in section 4 the experimental results.
In section 5 we show that for a Langevin equation the JE gives the
exact value of the free energy difference for any path. In section
6 we discuss the limits of applicability of JE and CR and we
conclude.

\section{The Jarzynski equality and the Crooks relation}
In 1997 \cite{jarzynski1} Jarzynski derived an equality which
relates the free energy difference of a system in contact with a
heat reservoir to the pdf of the work performed on the system to
drive it from $A$ to $B$ along any path $\ga$ in the system parameter space.
\subsection{The Jarzynski equality}
Specifically, when $\lam$ is varied from time $t = 0$ to $t =t_s$,
Jarzynski defines for one realization of the ``switching process''
from $A$ to $B$ the work performed on the system as
\begin{equation}
    W = \int_{0}^{t_s} \dot{\lambda}\, \frac{\partial H_{\lam} [z(t)]}
    {\partial \lam} \d t,
    \label{work}
\end{equation}
\noindent where $z$ denotes the phase-space point of the system
and $H_{\lam}$ its $\lam$-parametrized Hamiltonian (see also
\cite{landau_stat} and section 2.4). One can consider an ensemble
of realizations of this ``switching process'' with initial
conditions all starting in the same initial equilibrium state.
Then $W$ may be computed for each trajectory in the ensemble. The
JE states that \cite{jarzynski1}

\begin{equation}
    \Delta F = -\frac{1}{\beta} \ln \langle \exp{[-\beta W]} \rangle,
    \label{JE}
\end{equation}

\noindent where $\langle{\cdot}\rangle$ denotes the ensemble
average, $\beta^{-1} = k_B T$ with $k_B$ the Boltzmann constant
and $T$ the temperature. In other words $\langle \exp{[-\beta
W_{\mathrm{diss}}]} \rangle = 1$, since we can always write $W =
\Delta F+ W_{\mathrm{diss}}$ where $W_{\mathrm{diss}}$ is the
dissipated work. Thus it is easy to see that there must exist some
microscopic trajectories such that $W_{\textrm{diss}} \leq 0$.
Moreover, the inequality $\langle \exp{x} \rangle \geq
\exp{\langle x \rangle}$ allows us to recover the second
principle, namely $\langle W_{\textrm{diss}} \rangle \geq 0$, i.e.
$\langle W \rangle \geq \Delta F$. From an experimental point of
view the JE is quite useful because there is no restriction on the
choice of the path $\ga$ and it overcomes the above mentioned
experimental difficulties.

As we have already mentioned in the introduction, many proofs of
the JE have been done. The simplest way to understand JE is to
consider the perturbation theory \cite{landau_stat,peierls}. This
kind of approach has been  used by Landau \cite{landau_stat} but
he considered only Gaussian distributions and he stopped the
development to the second order. This technique  can be
generalized to get JE. Therefore we follow ref.\cite{landau_stat}
and we write the energy $H_B(p,q)$ of the system in contact with a
heath bath at temperature $T$ as:

\begin{equation}
H_B(p,q)=H_A(p,q)+ V(p,q,\lambda(t_s))
        \label{Energy}
\end{equation}

\noindent where $V$ is the energy introduced by the external
driving $\lambda$ such that $V(p,q,\lambda(0))=0$. By definition
the $\Delta F$ is given by

\begin{equation}
F_B=F_A-\beta^{-1} \ln \int \rho_A \ \exp{[-\beta V]} d \Gamma
\label{partition}
\end{equation}

\noindent where $\rho_A(p,q)=\exp{[\beta (F_A-H_A(p,q))]}$ is the
equilibrium Gibbs distribution in $A$, that is the unperturbed
state. Therefore
\begin{equation}
F_B=F_A-\beta^{-1} \ln < \exp{[-\beta V]} >  \label{DeltaF}.
\end{equation}
As ${dH\over dt}={\partial H \over \partial t}$
\cite{goldstein_mech} and  $H$ depends explicitly on time only by
mean of  $\lambda$ then

\begin{equation}
V=H_B-H_A=\int_0^{t_s} {dH\over  dt} dt =\int_0^{t_s} {\partial H
\over
\partial \lambda} {d\lambda\over dt} dt \label{eqV}.
\end{equation}

Comparing eq.\ref{eqV} to eq.\ref{work} we see that $V=W$ and
eq.\ref{DeltaF} is equivalent to eq.\ref{JE}.
\subsection{The Crooks relation}
In our experiment we can also check the CR which is somehow related
to the JE and which gives useful and complementary information on
the dissipated work. Crooks considers the forward work
$W_{\mathrm{f}}$ to drive the system from $A$ to $B$ and the
backward work $W_{\mathrm{b}}$ to drive it from $B$ to $A$. If the
work pdfs during the forward and backward processes are
$\mathrm{P}_{\mathrm{f}}(W)$ and $\mathrm{P}_{\mathrm{b}}(W)$, one
has \cite{crooks1, jarzynski2}
\begin{equation}
    \frac{\mathrm{P}_{\mathrm{f}}(W)}{\mathrm{P}_{\mathrm{b}}(-W)}
    = \exp{(\beta [W-\Delta F])}
    = \exp{[\beta W_{\mathrm{diss}}]}
    \label{crooks}.
\end{equation}
A simple calculation from Eq.\,(\ref{crooks}) leads to
Eq.\,(\ref{JE}). However, from an experimental point of view this
relation is extremely useful because one immediately sees that the
crossing point of the two pdfs, that is the point where
$\mathrm{P}_{\mathrm{f}}(W) = \mathrm{P}_{\mathrm{b}}(-W)$, is
precisely $\Delta F$. Thus one has another mean to check the
computed free energy by looking at the pdfs crossing point
$W_{\mathrm{\times}}$.
\subsection{The Gaussian case}
Let us examine in some detail the Gaussian case, that is
$\mathrm{P}(W) \propto \exp{\Bigl(-\frac{[W - \langle W
\rangle]^2}{2 \sigma_W^2}}\Bigr)$. In this case the JE leads to
\begin{equation}
    \Delta F = \langle W \rangle - \frac{\beta \sigma_W^2}{2},
    \label{eqgausF}
\end{equation}
i.e. $\langle W_{\textrm{diss}} \rangle = \frac{\beta
\sigma_W^2}{2} > 0$. It is interesting to notice that Landau
\cite{landau_stat}  derived  eq.\ref{eqgausF}, which can also be
viewed as a consequence of the linear response theory as it has
been shown  by Hermans \cite{Hermans}.

It is easy to see from Eq.\,(\ref{crooks}) that if
$\mathrm{P}_{\mathrm{f}}(W)$ and $\mathrm{P}_{\mathrm{b}}(-W)$ are
Gaussian, then
\begin{equation}
    \Delta F = \frac{\langle W \rangle_{\mathrm{f}} - \langle W \rangle_{\mathrm{b}}}{2},
\end{equation}
and
\begin{equation}
    \qquad \beta \sigma_W^2 = \langle W \rangle_{\mathrm{f}} +
    \langle W \rangle_{\mathrm{b}} = 2\,\langle W_{\mathrm{diss}} \rangle.
\end{equation}
 Thus in the case of Gaussian statistics $\Delta F$ and
$W_{\mathrm{diss}}$ can be computed by using just the mean values and the
variance of the work $W$.
\subsection{The classical work and the $\Delta F$ computed by the JE}
Before describing the experiment, we want to discuss several important points.
The first is the definition of the work given in Eq.\,(\ref{work}), which is not
the classical one. Let us consider, for example, that $\lambda$ is a mechanical
torque $M$ applied to a mechanical system $\Xi$, and
$-\partial H_{\lam} / \partial \lam$ the associated angular displacement
$\theta$. Then, from Eq.\,(\ref{work}), one has
\begin{equation}
    W = -\int_{0}^{t_s} \dot{M} \theta \d t = -\Bigl[M \theta \Bigr]_{0}^{t_s} - W^{\mathrm{cl}},
\end{equation}
where
\begin{equation}
     W^{\mathrm{cl}} = -\int_{0}^{t_s} M \dot{\theta} \d t
     \label{classicalwork}
\end{equation}
is the classical work (we define the classical work with minus
sign to respect the standard convention of thermodynamics). Thus
$W$ and $W^{\mathrm{cl}}$ are related but they are not exactly the
same and we will show that this makes an important difference in
the fluctuations of these two quantities. This difference between
the $W$ and $W^{cl}$, has been already pointed out in
ref.\cite{Hummer}.

 The second important  point that we want to discuss concerns the $\Delta F$
computed by the JE in the case of a driven system, composed by
$\Xi$ plus the external driving. The total free energy difference
is
\begin{equation}
    \Delta F = \Delta F_0 - \Bigl[M \theta \Bigr]_A^B = \Delta F_0 - \Phi,
    \label{eqdeltaF}
\end{equation}
where $\Delta F_0$ is the free energy of $\Xi$ and $\Phi = \Bigl[M
\theta \Bigr]_A^B$ the energy difference of the forcing. The JE
computes the $\Delta F$ of the driven system and not that of the
system alone which is $\Delta F_0$. This is an important
observation in view of all applications where an external
parameter is added to $\Xi$ in order to measure $\Delta F_0$
\cite{bustamante}. Finally we point out that, in an isothermal
process, $\Delta F_0$ can be easily computed, without using the JE
and the CR, if $W^{\mathrm{cl}}$ is Gaussian distributed with
variance $\sigma_{W^{\mathrm{cl}}}^2$. (We make the reasonable
assumption that forward and backward work variances are equal.)
Indeed the crossing point $W_{\times}^{\mathrm{cl}}$ of the two
Gaussian pdfs $\mathrm{P}_{\mathrm{f}} (W^{\mathrm{cl}})$ and
$\mathrm{P}_{\mathrm{b}} (-W^{\mathrm{cl}})$ is
\begin{equation}
    W_{\times}^{\mathrm{cl}} = \frac{\langle W^{\mathrm{cl}} \rangle_{\mathrm{f}}
    - \langle W^{\mathrm{cl}} \rangle_{\mathrm{b}}}{2},
\end{equation}
which by definition is just $-\Delta F_0$, i.e.
$W_{\times}^{\mathrm{cl}}= -\Delta F_0$. Furthermore the
dissipated work can be obtained from
\begin{equation}
    \langle W_{\mathrm{diss}} \rangle \ = - { \langle
    W^{\mathrm{cl}} \rangle_{\mathrm{f}} + \langle W^{\mathrm{cl}}
    \rangle_{\mathrm{b}} \over 2},
\end{equation}
by definition. It should be noted that the equality
$2\,\langle W_{\mathrm{diss}} \rangle = \beta \sigma_{W^{\mathrm{cl}}}^2$
does not hold in the case of the classical work.
\begin{figure}
    \begin{center}
    \includegraphics[width=5cm, angle=0]{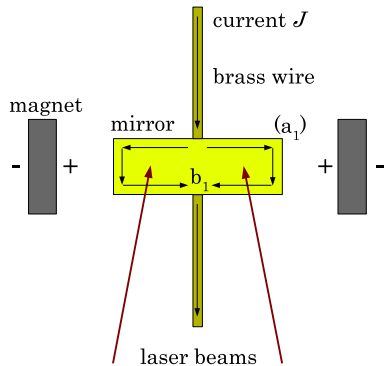}
    \caption{Schematic drawing of the oscillator}
    \label{setup_and_fdt_check}
    \end{center}
\end{figure}

\begin{figure}
    \begin{center}
    {\hspace{3.5cm} (i) \hspace{3.5cm} (ii)  \hspace{3.5cm} (iii)} \\
    \includegraphics[width=4.7cm, angle=0]{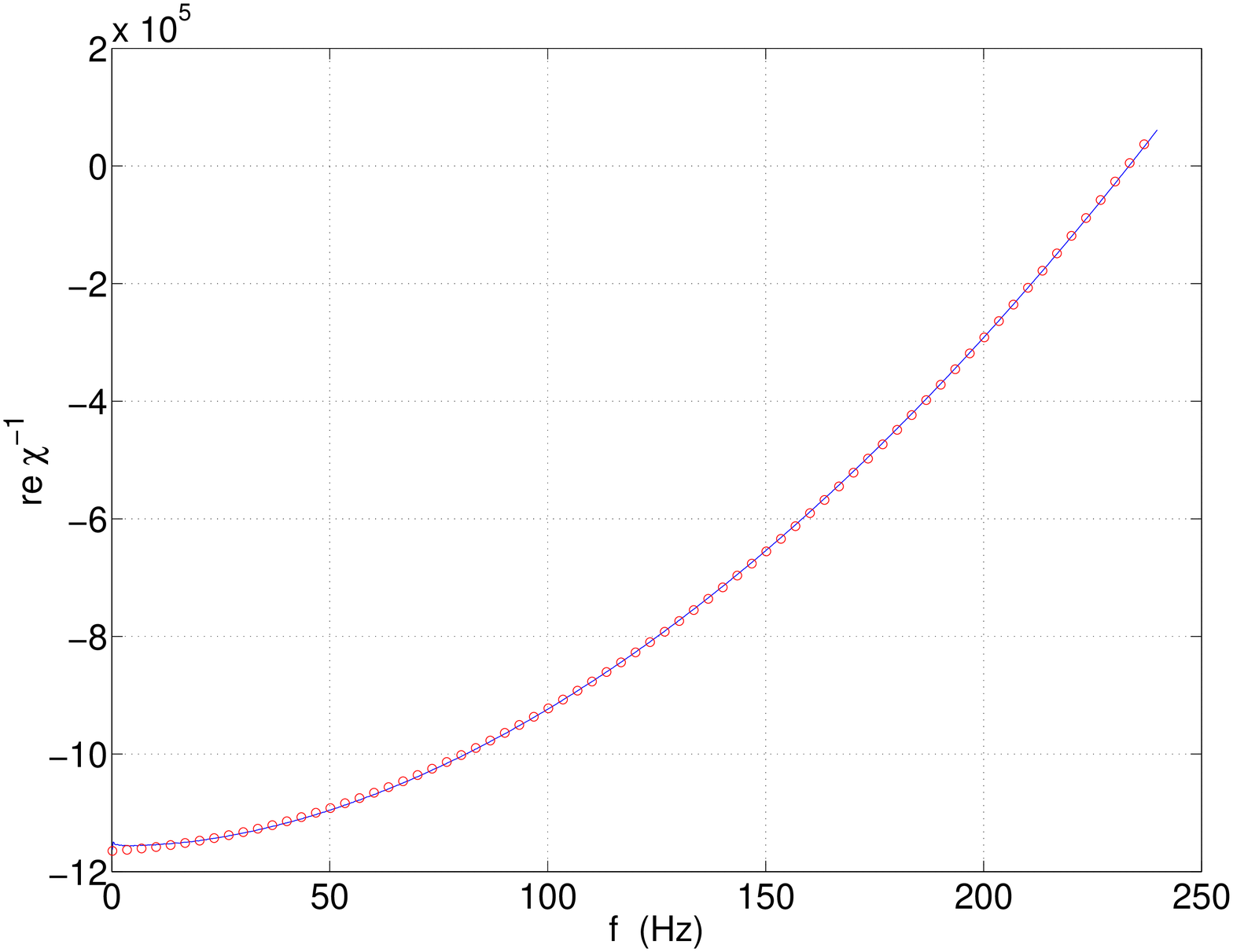}
    \includegraphics[width=4.7cm, angle=0]{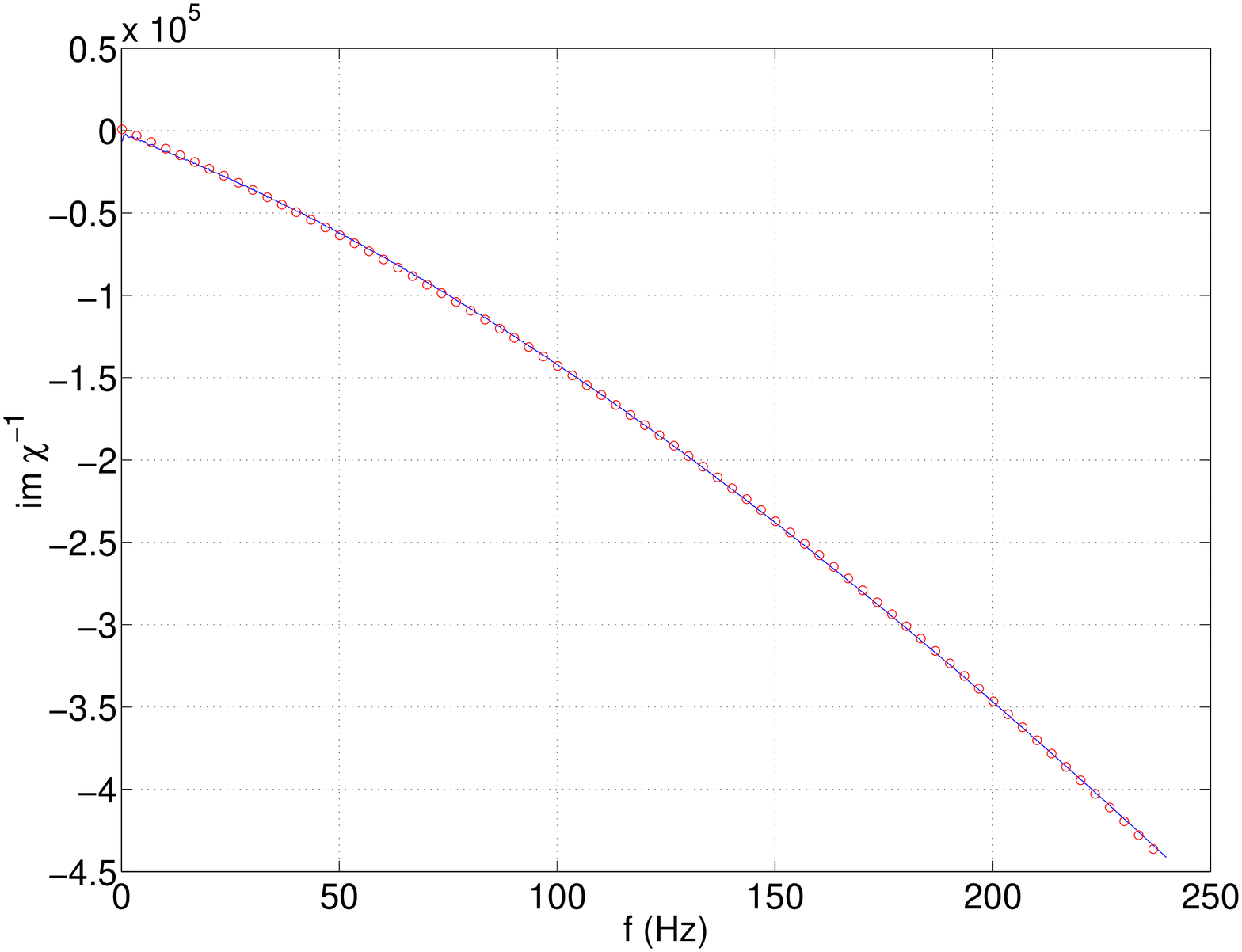}
    \includegraphics[width=5 cm, angle=0]{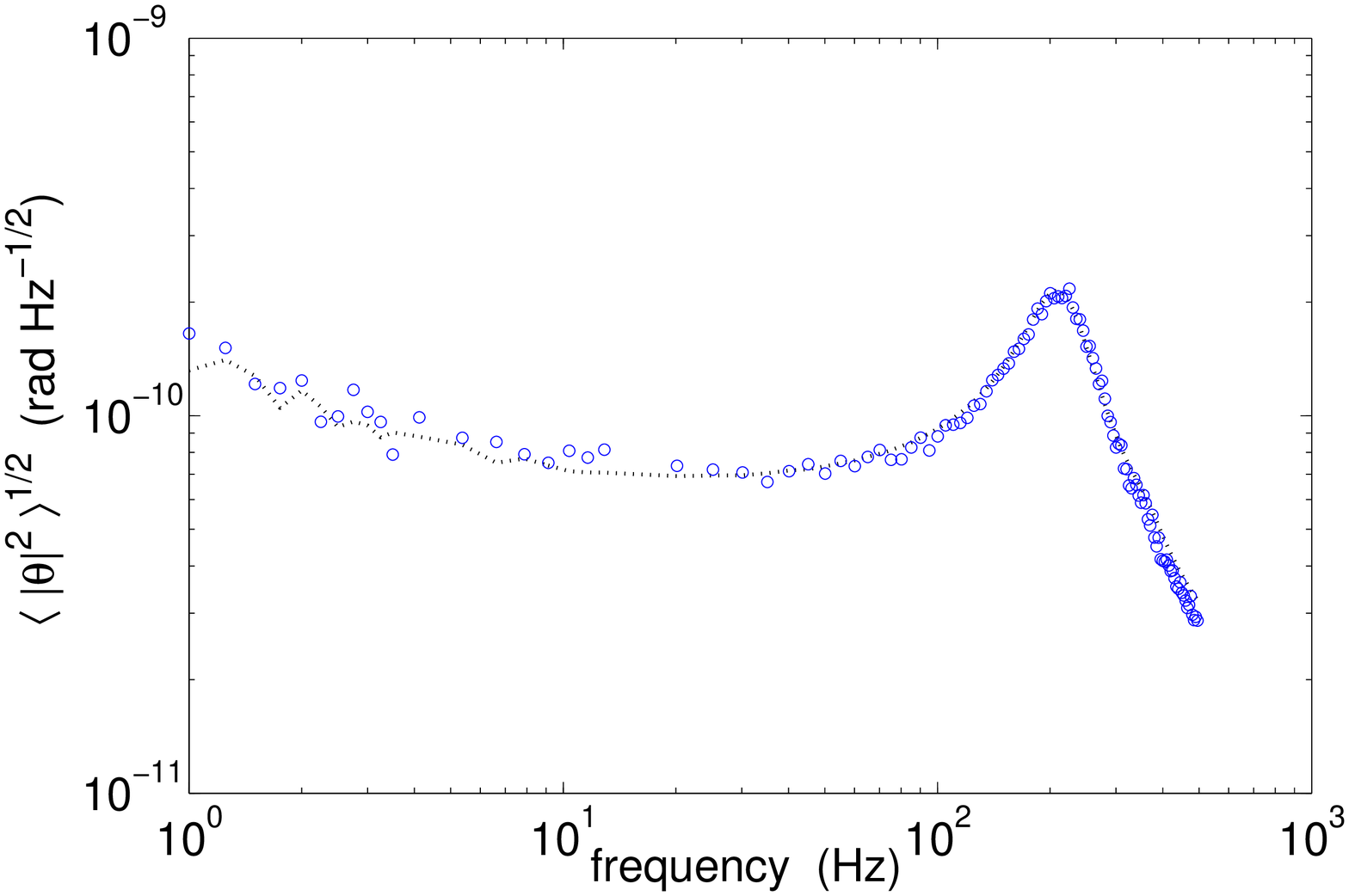}
    \caption{ FDT check in oil. i) Real part $1/\chi$. ii)
    Imaginary part of $1/\chi$. iii) Noise spectrum
    the circles $(\circ)$ correspond to the direct measurement of the noise and
    the dashed curve is $\sqrt{\langle { \vert \hat{\theta} \vert }^2 \rangle}$
    computed by inserting the measured response function $\chi$ in the FDT, Eq.\,(\ref{fdt})}
    \label{fdt_check}
    \end{center}
\end{figure}

\section{Experimental setup}
\subsection{The torsion pendulum}
 To study the JE and the CR we measure the
out-of-equilibrium fluctuations of a macroscopic mechanical
torsion pendulum made of a brass wire, whose damping is given
either by the viscoelasticity of the torsion wire or by the
viscosity of a surrounding fluid. This system is enclosed in a
cell which can be filled with a viscous fluid, which acts as a
heat bath. A brass wire of length $10 \textrm{ mm}$, width $0.75
\textrm{ mm}$, thickness $50 \textrm{ } \mu \textrm{m}$, mass
$5.91 \times 10^{-3} \textrm{ g}$, is clamped at both ends, hence
its elastic torsional stiffness is $C = 7.50 \times 10^{-4}
\textrm{ N\,m}\mathrm{\,rad^{-1}}$. A small mirror of effective
mass $4.02 \times 10^{-2} \textrm{ g}$, length $2.25 \textrm{
mm}$, width $b_1 = 7 \textrm{ mm}$, thickness $a_1 = 1.04 \textrm{
mm}$, is glued in the middle of the wire, see
Fig.\,\ref{setup_and_fdt_check}(i), so that the moment of inertia
of the wire plus the mirror in vacuum is $I = 1.79 \times 10^{-10}
\textrm{ kg}\mathrm{\,m^2}$ (whose main contribution comes from
the mirror). Thus the resonant frequency of the pendulum in vacuum
is $f_0 = 326.25 \textrm{ Hz}$. When the cell is filled with a
viscous fluid, the total moment of inertia is $I_{\mathrm{eff}} =
I + I_{\mathrm{fluid}}$, where $I_{\mathrm{fluid}}$ is the extra
moment of inertia given by the fluid displaced by the mirror
\cite{lamb}. Specifically, for the oil used in the experiment
(which is a mineral oil of optical index  $n = 1.65$, viscosity $ 
121.3 \textrm{ mPa\,s}$ and density $\rho =
0.9\,\rho_{\mathrm{water}}$ at $T = 21.3 \,^{\circ}\mathrm{C}$)
the resonant frequency becomes $f_0 = 213 \textrm{ Hz}$. To apply
an external torque $M$ to the torsion pendulum, a small electric
coil connected to the brass wire is glued in the back of the
mirror. Two fixed magnets on the cell facing each other with
opposite poles generate a static magnetic field. We apply a torque
by varying a very small current $J$ flowing through the electric
coil, hence  $M= Amp \  J$, where $Amp$ is an amplification factor
which depends on the size of the coil and of the distance of the
magnets. This factor, which can be measured independently (see
ref.\cite{Bellon},  is the largest source of error of our
measurement. It is known with $3\%$ of accuracy.
The measurement of the angular displacement of the mirror $\theta$
is done using a Nomarski interferometer \cite{nomarski,
optics_com} whose noise is about $6.25 \times 10^{-12} \textrm{
rad}/\sqrt{\textrm{Hz}}$, which is two orders of magnitude smaller
than the oscillator thermal fluctuations. An optical window lets
the laser beams  to go inside and outside cell. Much care has been
taken in order to isolate the apparatus from the external
mechanical and acoustic noise, see \cite{rsi} for details.

\subsection{Equation of motion}
The motion of the torsion pendulum can be assimilated to that of a
driven harmonic oscillator damped by the viscoelasticity of the
torsion wire and the viscosity of the surrounding fluid, whose
motion equation reads in the temporal domain
\begin{equation}
    I_{\mathrm{eff}}\,\ddot{\theta} + \int_{-\infty}^{t} G(t-t')\, \dot{\theta}(t') \d t' + C \theta = M,
    \label{eqofmotion}
\end{equation}
where $G$ is the memory kernel, which in the simplest case of a
viscous damping is  $G(t-t')=\nu \delta(t-t')$. In Fourier space
(in the frequency range of our interest) this equation takes the
simple form
\begin{eqnarray}
    [- I_{\mathrm{eff}}\,{\om}^2 + \hat{C}]\, \hat{\theta} = \hat{M},
\end{eqnarray}
where $\hat{\cdot}$ denotes the Fourier transform and $\hat{C} = C
+ i [C' + \om \nu ]$ is the complex frequency-dependent elastic
stiffness of the system. $C'$ and $\nu$ are  the viscoelastic and
viscous components of the damping term. The response function of
the system $\hat{\chi} = \hat{\theta} / \hat{M}$ can be measured
by applying a torque with a white spectrum. When $M = 0$, the
amplitude of the thermal vibrations of the oscillator is related
to its response function via the fluctuation-dissipation theorem
(FDT) \cite{landau_stat}. Therefore, the thermal fluctuation power
spectral density (psd) of the torsion pendulum reads for positive
frequencies
\begin{equation}
    \langle { \vert \hat{\theta} \vert }^2 \rangle
    = \frac{4 k_B T}{\om} \, \mathrm{Im} \, \hat{\chi}
    =\frac{4 k_B T}{\om} \frac{C' + \om \, \nu}
    {{\lbrack -I_{\mathrm{eff}}\,{\om}^2 + C \rbrack}^2 + [C' + \om \, \nu]^2}.
    \label{fdt}
\end{equation}
\subsection{Calibration and accuracy of the measurements}
To calibrate the system we check whether FDT is satisfied by the
measurements.  We apply a current $J$ with a white spectrum. We
measure the transfer function $\hat{M}/\hat{\theta}=1/\hat{\chi}=
[- I_{\mathrm{eff}}\,{\om}^2 + C + i (C_1'' + \om \nu)]$. As an
example we plot in fig.\ref{fdt_check}i) the measured real part of
$1/\hat{\chi}$ and in ii) the imaginary part. The parabolic fit of
the real part and the polynomial fit of the imaginary part ($\nu$
may depend on frequency) allow us to estimate all the parameters
of the oscillator. The main source of error is given by the
amplitude of the applied torque which is known with about $3 \%$
accuracy. Once $\chi$ is known the power spectrum  of $\theta$
with $M=0$ can be computed from eq.\ref{fdt} We plot in
Fig.\,\ref{fdt_check}(iii) the measured thermal square root psd of
the oscillator. The measured noise spectrum [circles in
Fig.\,\ref{fdt_check}(iii)] is compared with the one estimated
[dotted line in Fig.\,\ref{fdt_check}(iii)] by inserting the
measured $\hat{\chi}$ in the FDT, Eq.\,(\ref{fdt}). The two
measurements are in perfect agreement and obviously the FDT is
fully satisfied because the system is at equilibrium in the state
$A$ where $M = 0$ (see below). Although this result is expected,
this test is very useful to show that the experimental apparatus
can measure with a good accuracy and resolution the thermal noise
of the macroscopic pendulum.

\begin{figure}
    \begin{center}
    \includegraphics[width=4.6cm, angle=0]{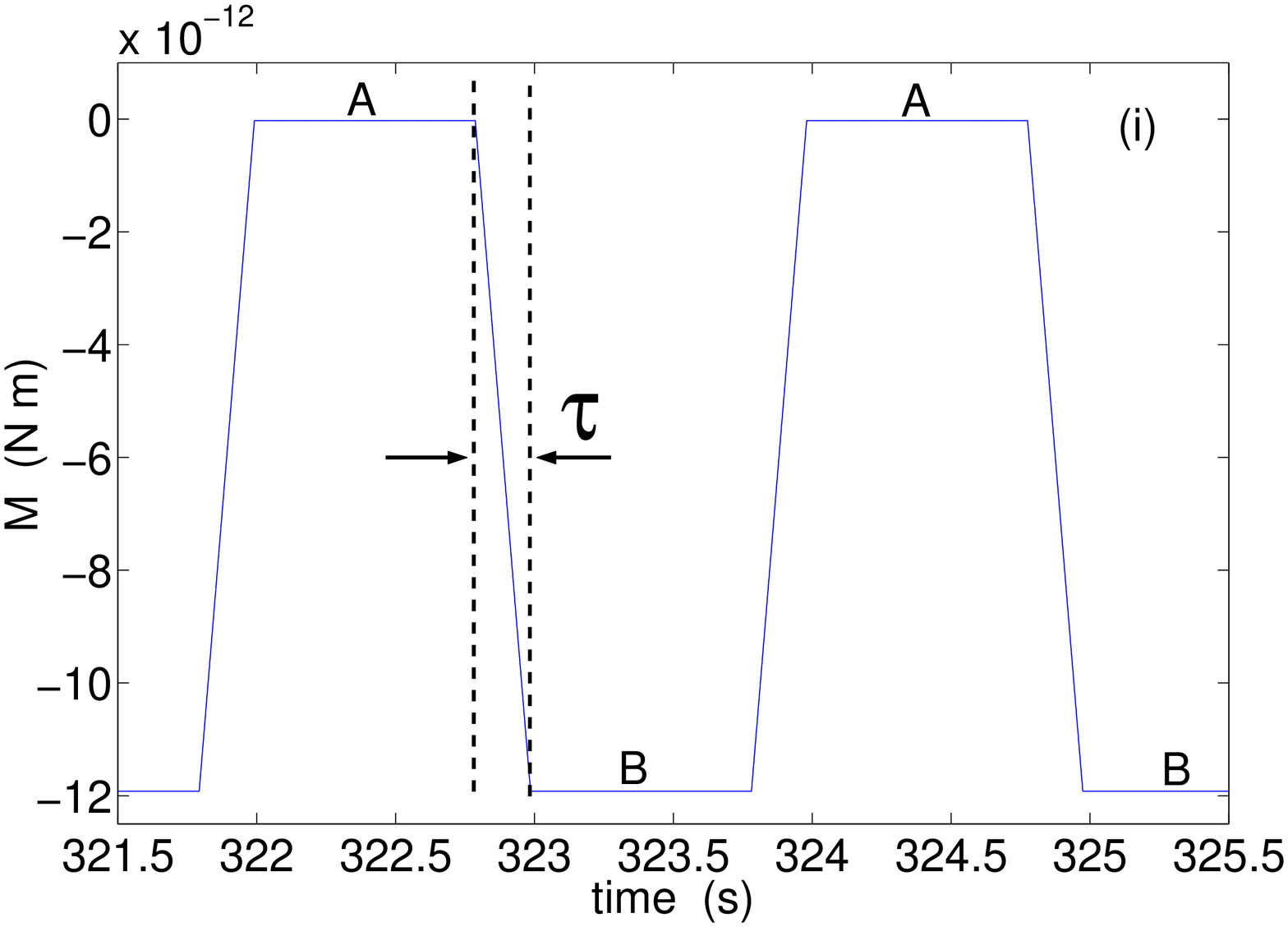} 
    \includegraphics[width=4.6cm, angle=0]{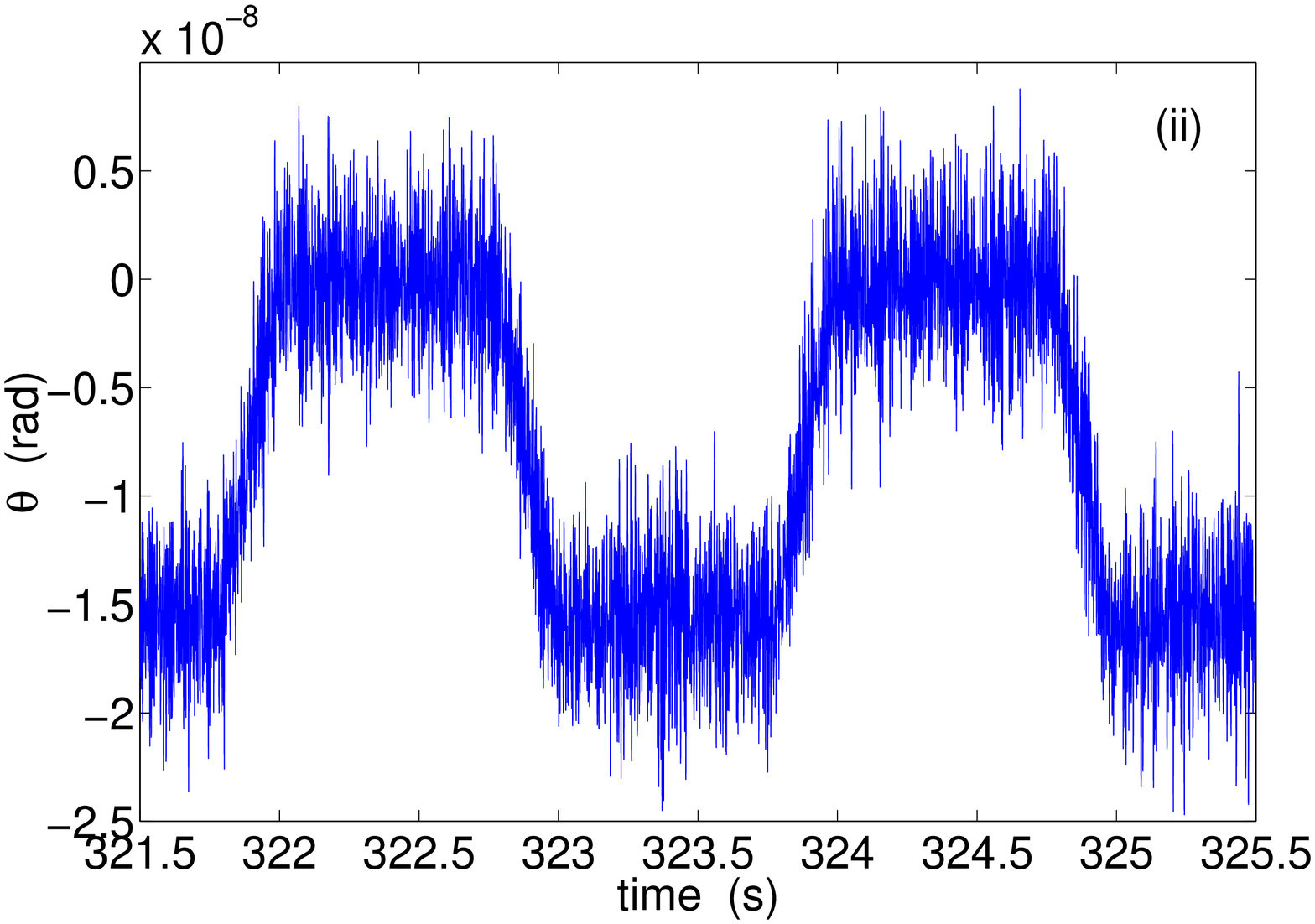}
    \includegraphics[width=4.6cm, angle=0]{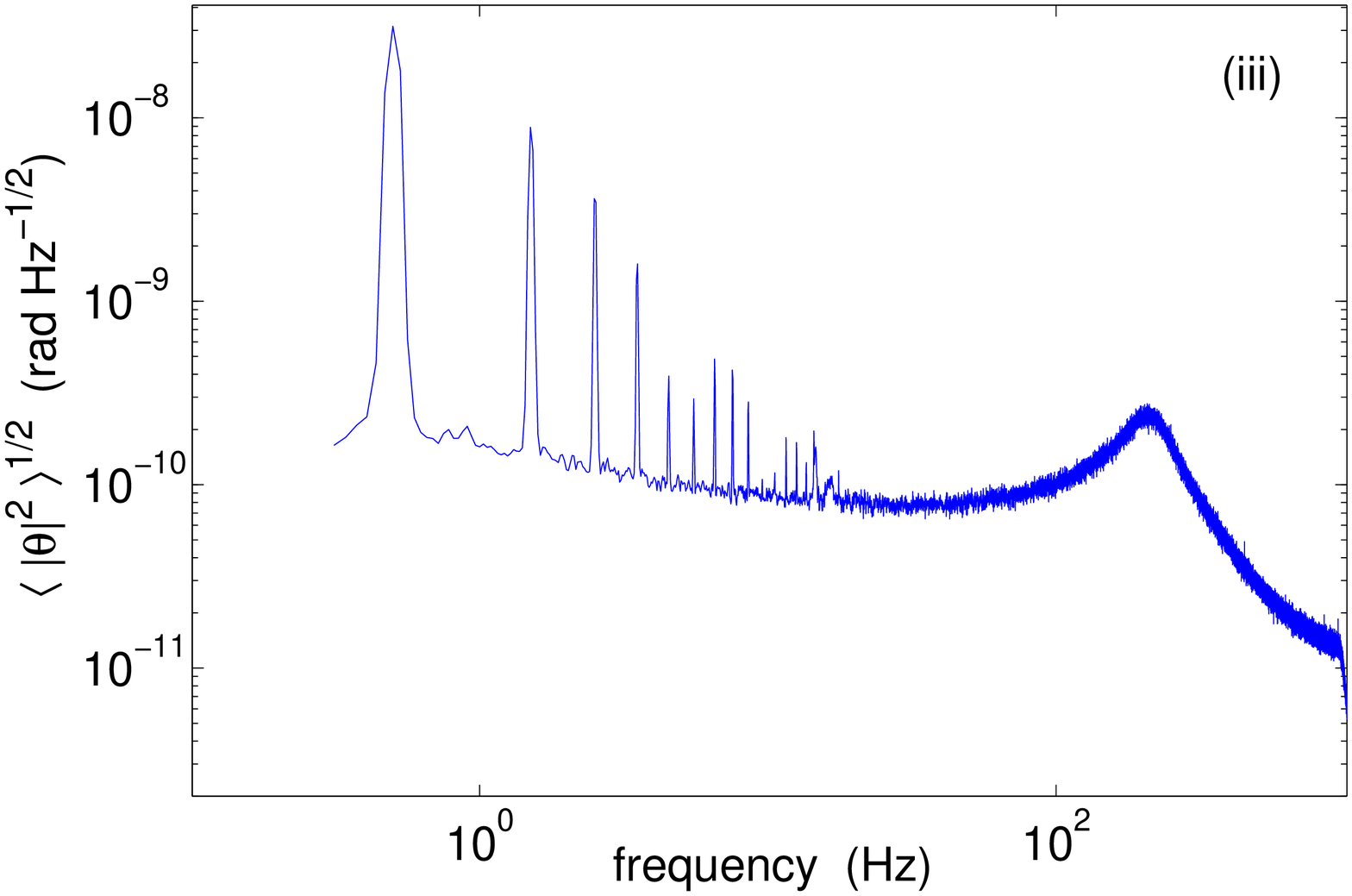}\\
    \includegraphics[width=4.6cm, angle=0]{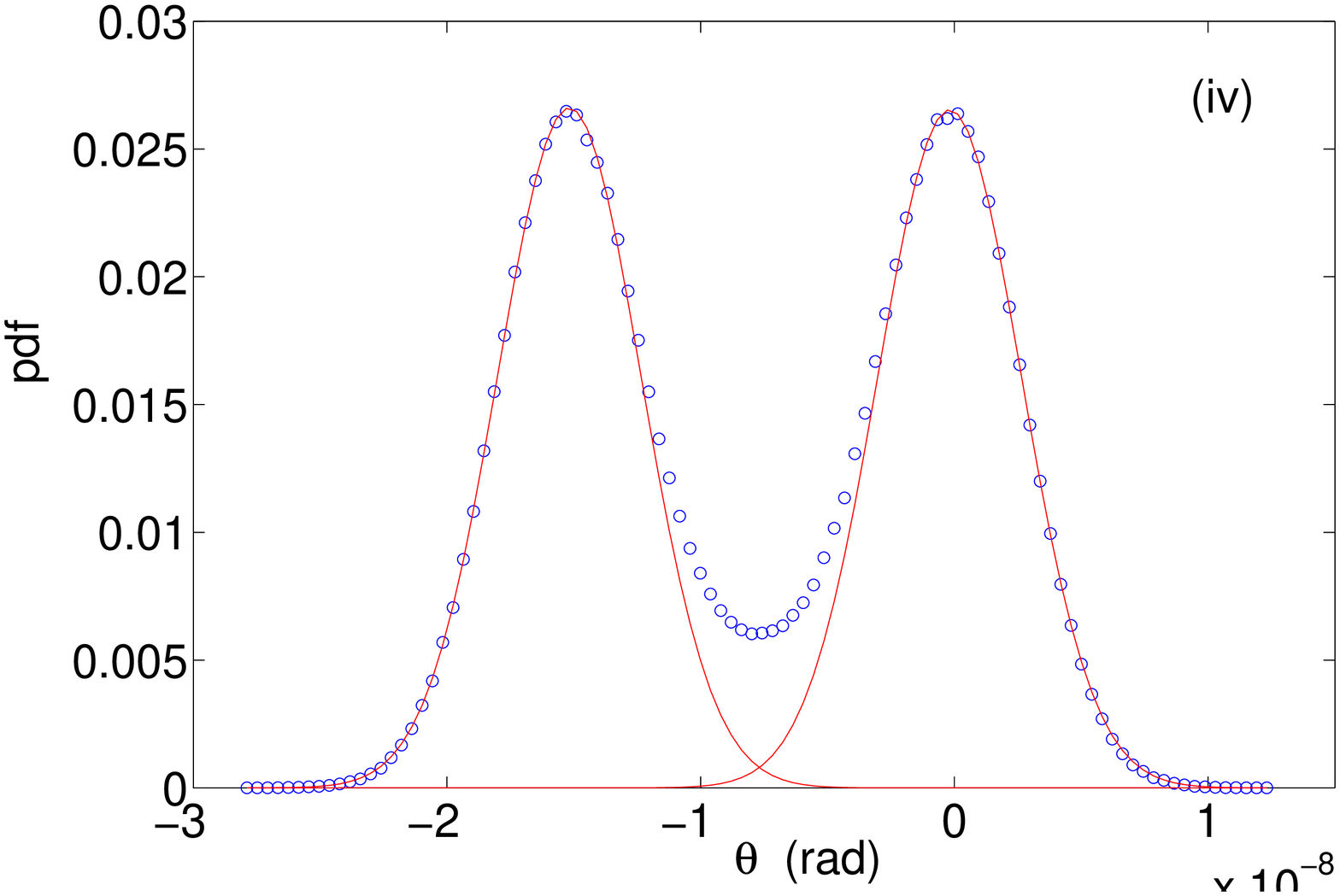}
    \includegraphics[width=4.6cm, angle=0]{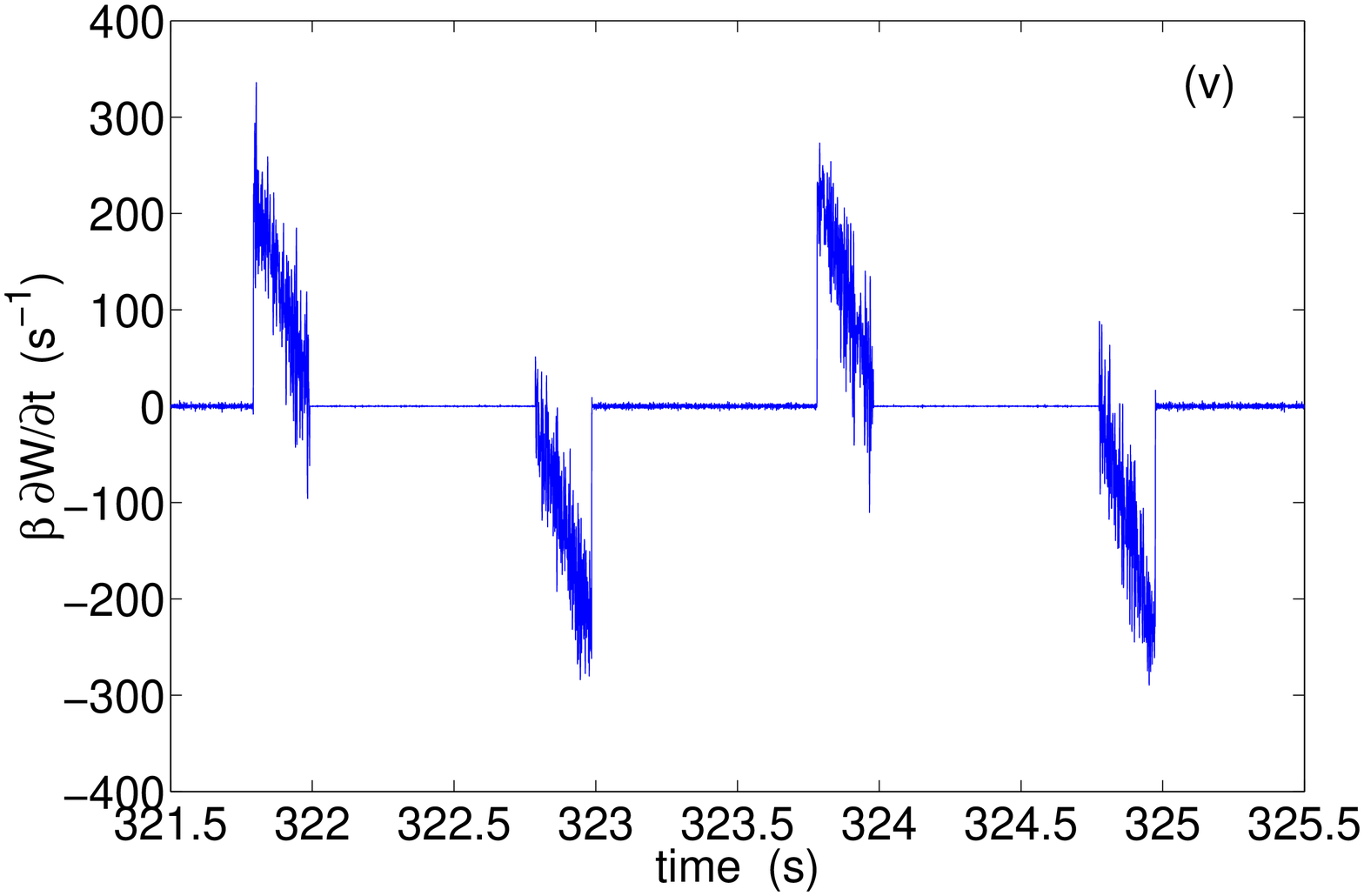}
    \includegraphics[width=4.6cm, angle=0]{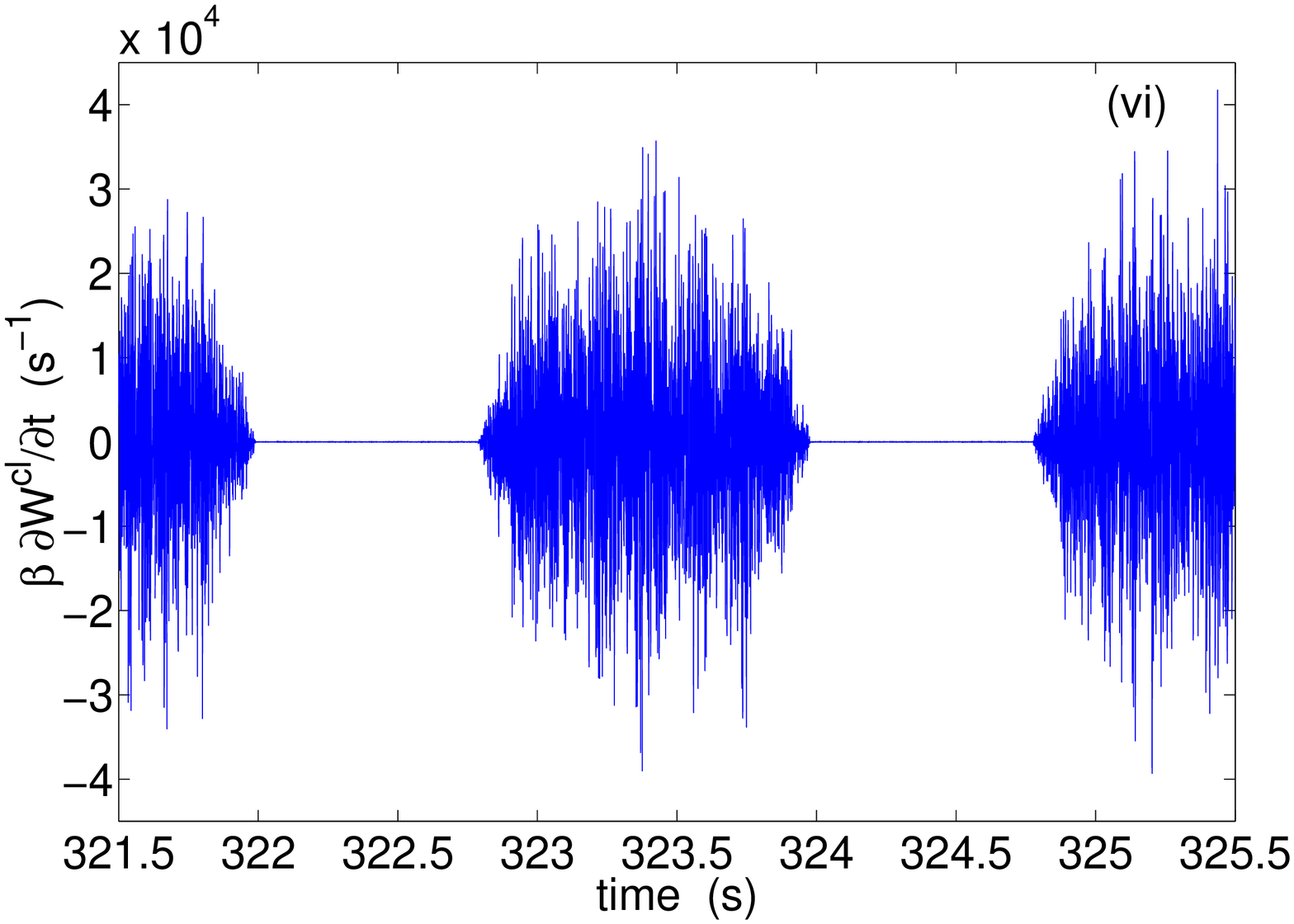}
    \caption{
    Oscillator immersed in oil [case (a)]:
    (i) Applied external torque,
    (ii) Induced angular displacement,
    (iii) its psd, 
    (iv) its pdf,
    (v) Injected power computed from the Jarzynski definition $\dot{W} = -\dot{M} \theta$,
    (vi) Injected power computed from the standard definition $\dot{W}^{\mathrm{cl}} = -M \dot{\theta}$}
    \label{driver_fluct_spec_displacement_displacement_pdf_power_class_power}
    \end{center}
\end{figure}

\begin{figure}[h!]
    \begin{center}
    \includegraphics[width=6cm, angle=0]{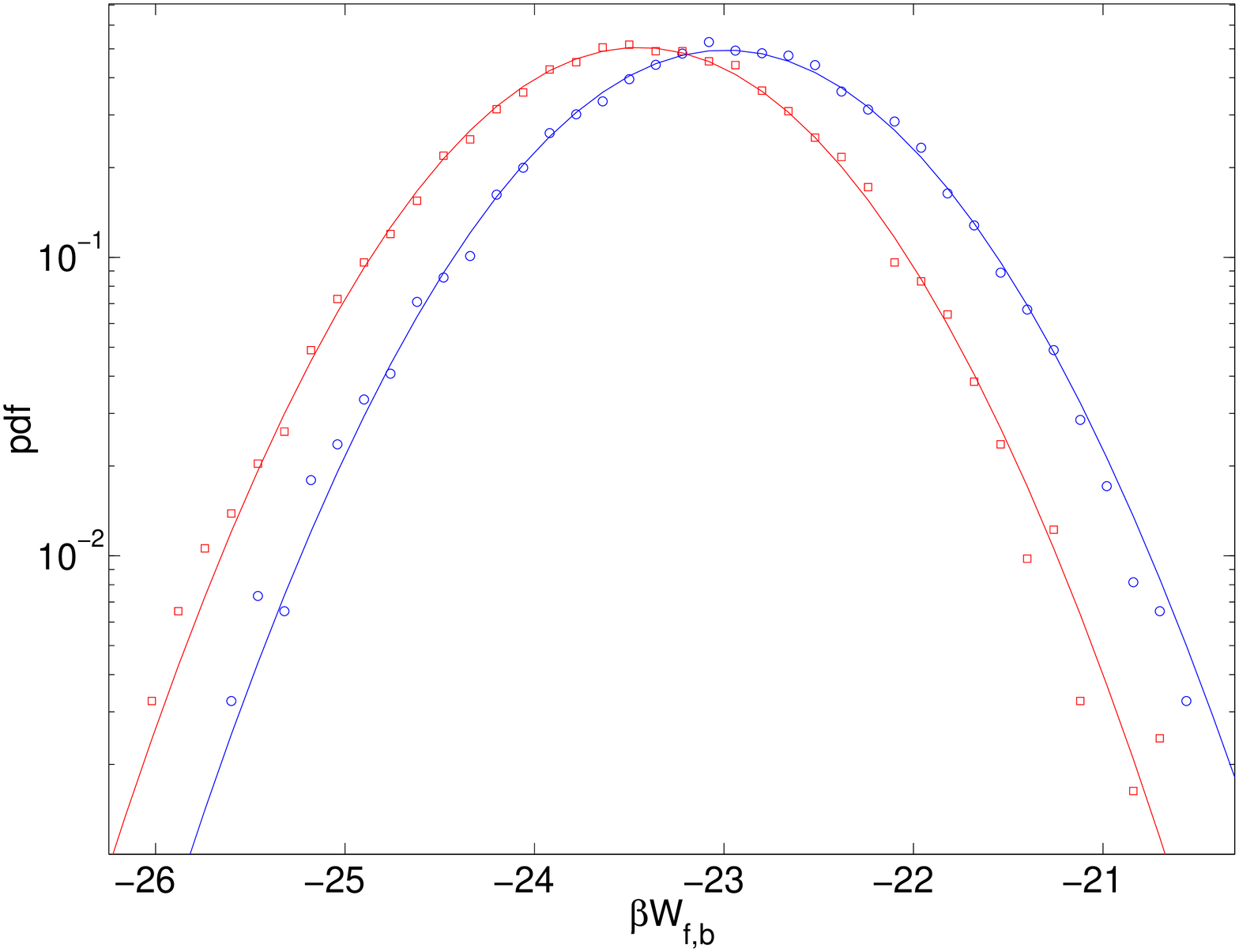}
    \includegraphics[width=6cm, angle=0]{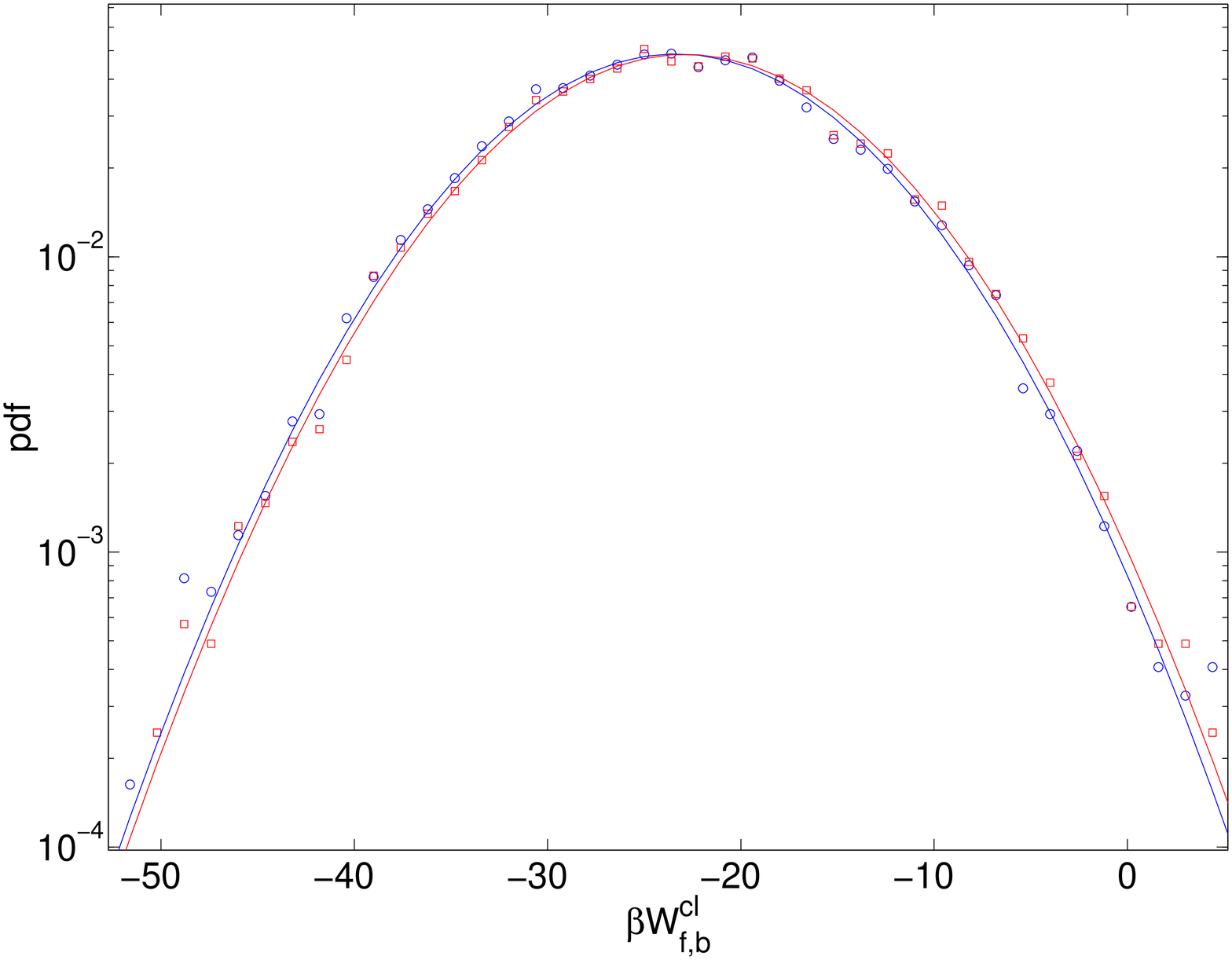}\\
    \includegraphics[width=6cm, angle=0]{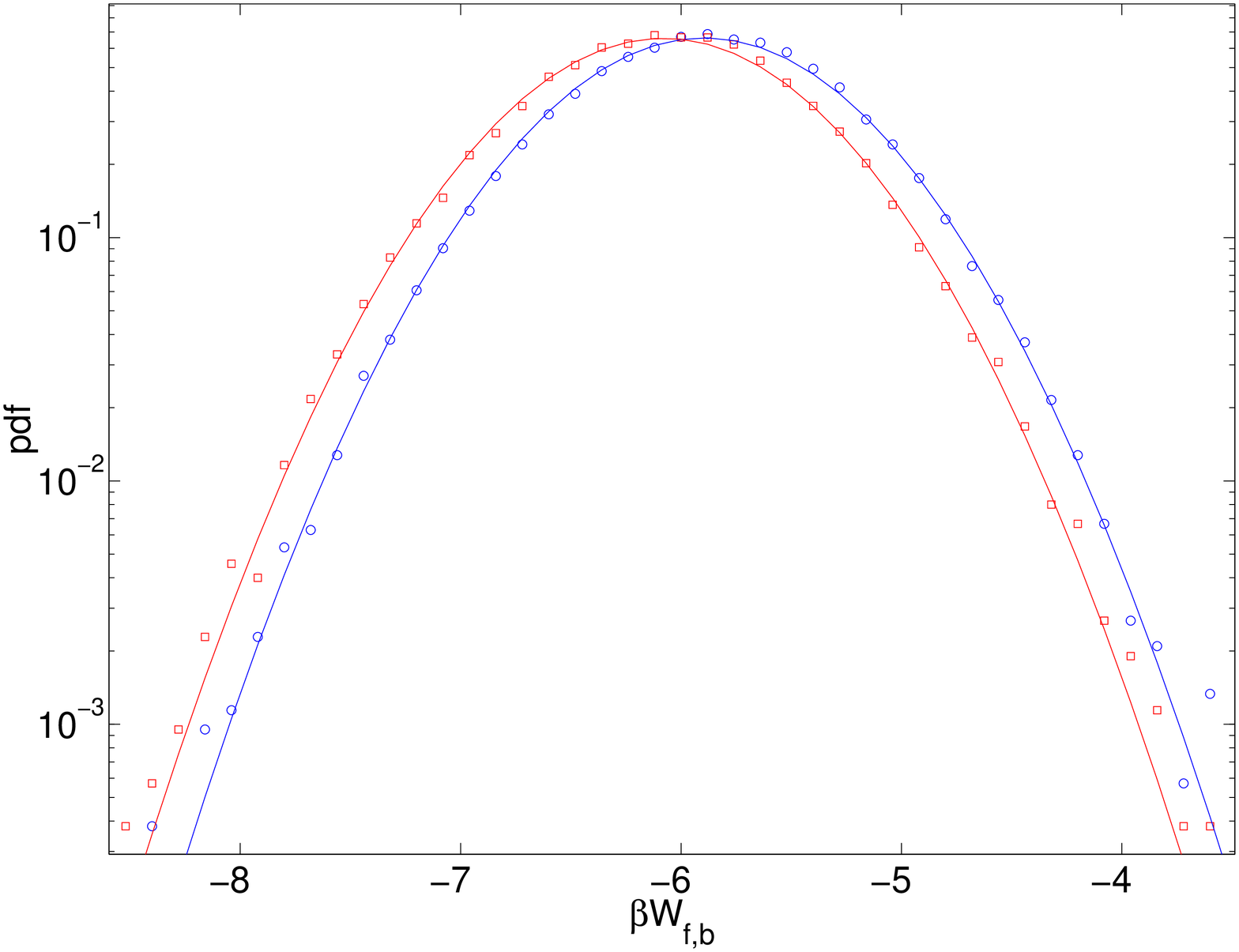}
    \includegraphics[width=6cm, angle=0]{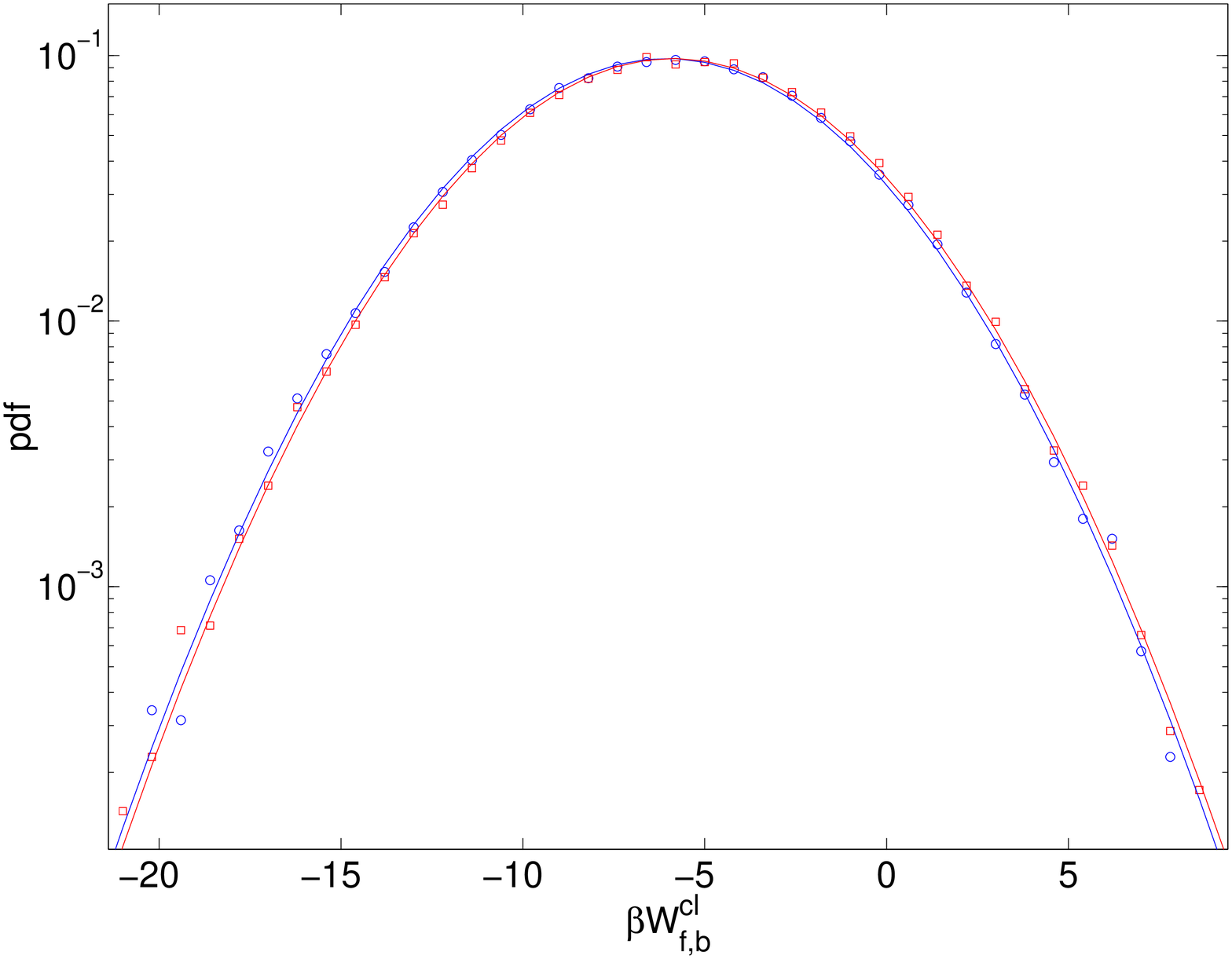}
    \caption{Case a): (i) $\mathrm{P}_{\mathrm{f}}(W)$ and $\mathrm{P}_{\mathrm{b}}(-W)$,
    (ii) $\mathrm{P}_{\mathrm{f}}(W^{\mathrm{cl}})$ and $\mathrm{P}_{\mathrm{b}}(-W^{\mathrm{cl}})$;
    Case c): (iii) $\mathrm{P}_{\mathrm{f}}(W)$ and $\mathrm{P}_{\mathrm{b}}(-W)$,
    (iv) $\mathrm{P}_{\mathrm{f}}(W^{\mathrm{cl}})$ and $\mathrm{P}_{\mathrm{b}}(-W^{\mathrm{cl}})$
    (experimental forward and backward pdfs are represented by $\circ$ and $\Box$ respectively,
    whereas the continuous lines are Gaussian fits)}
    \label{pdfs}
    \end{center}
\end{figure}
\section{Experimental results}
\begin{figure}[h!]
    \begin{center}
{\hspace{6cm} (i) \hspace{7cm} (ii)}\\
    \includegraphics[width=7cm, angle=0]{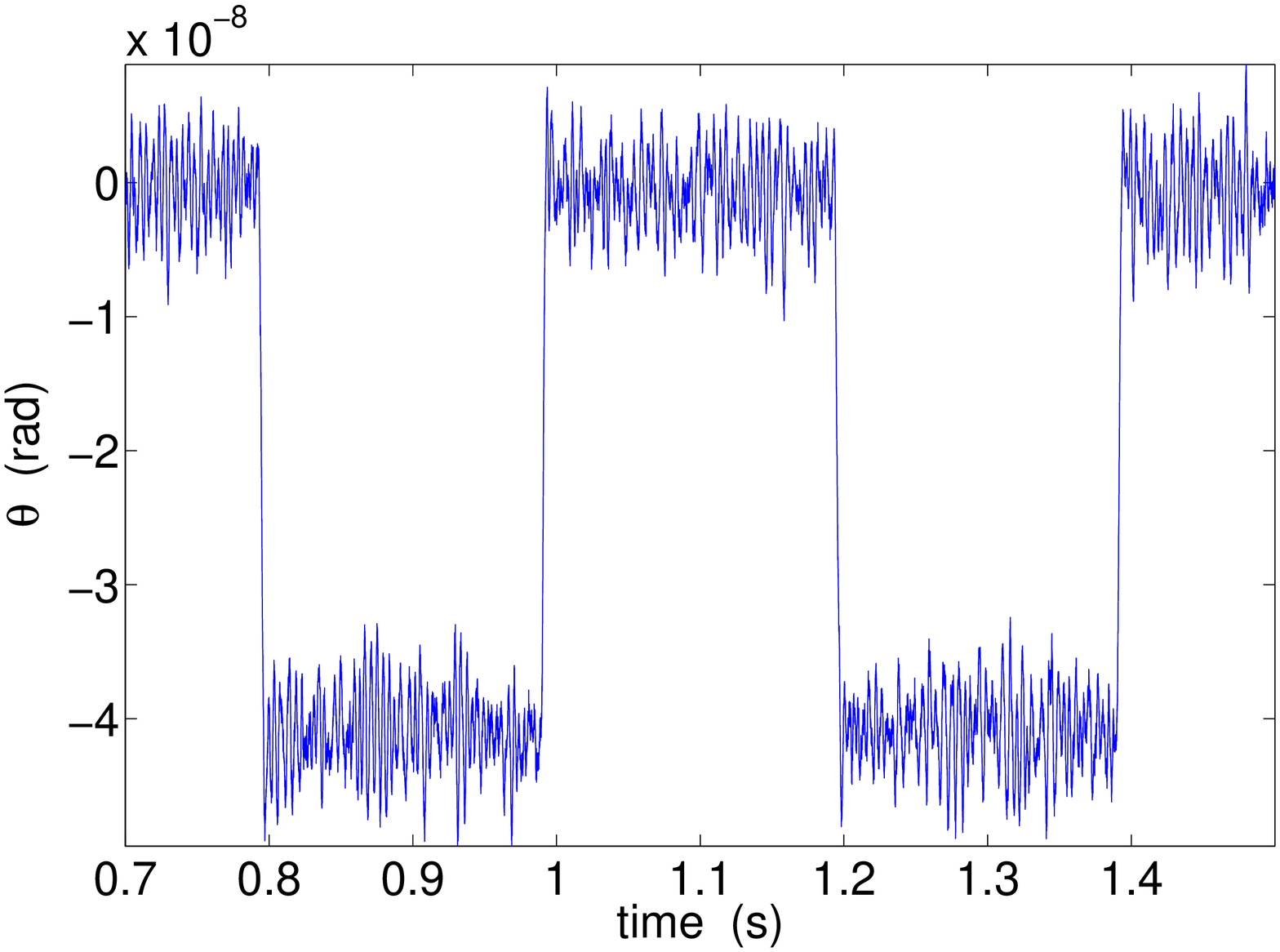}
    \includegraphics[width=7cm, angle=0]{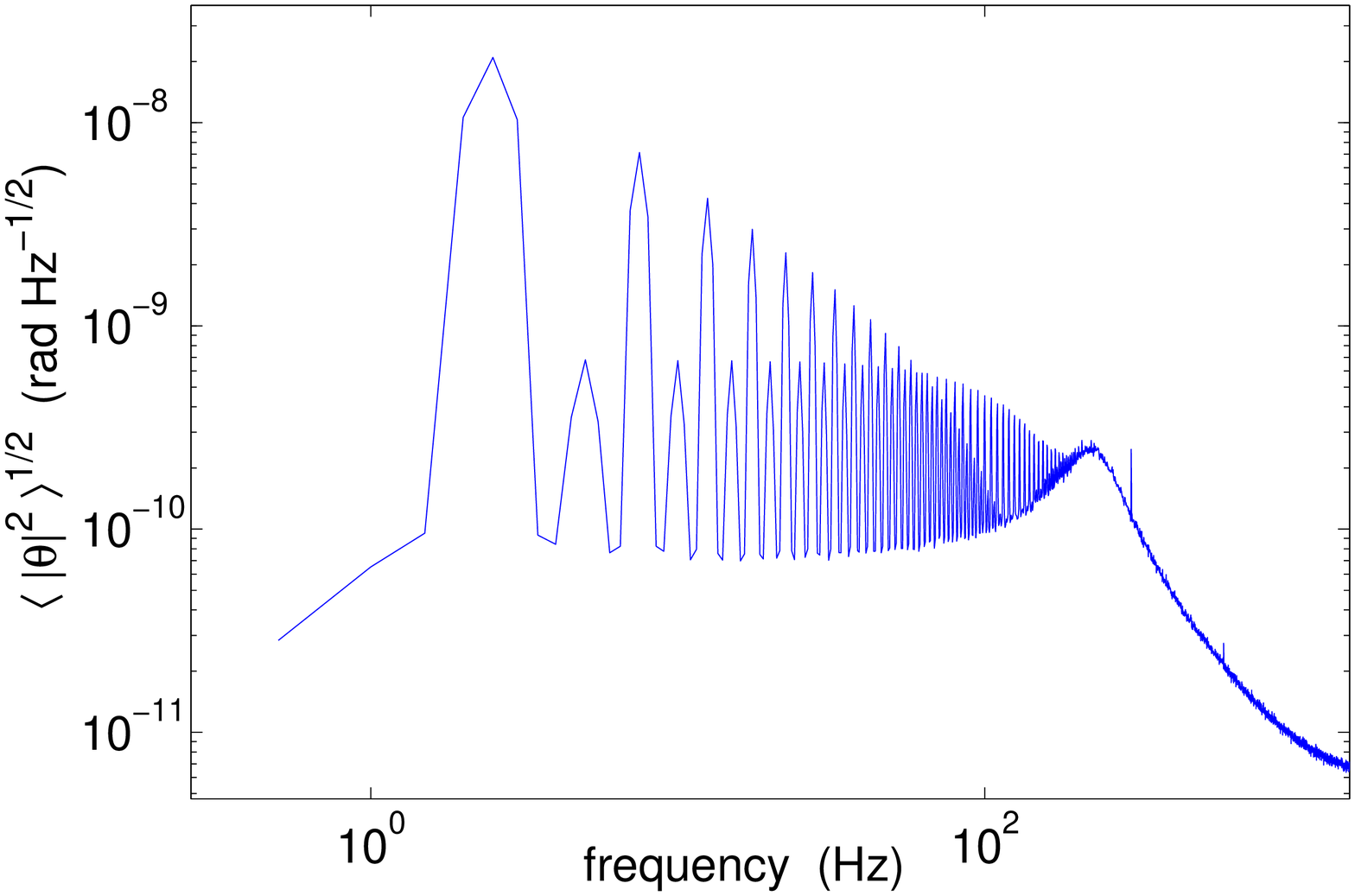}\\
    {\hspace{6cm} (iii) \hspace{7cm} (iv)}\\
    \includegraphics[width=7cm, angle=0]{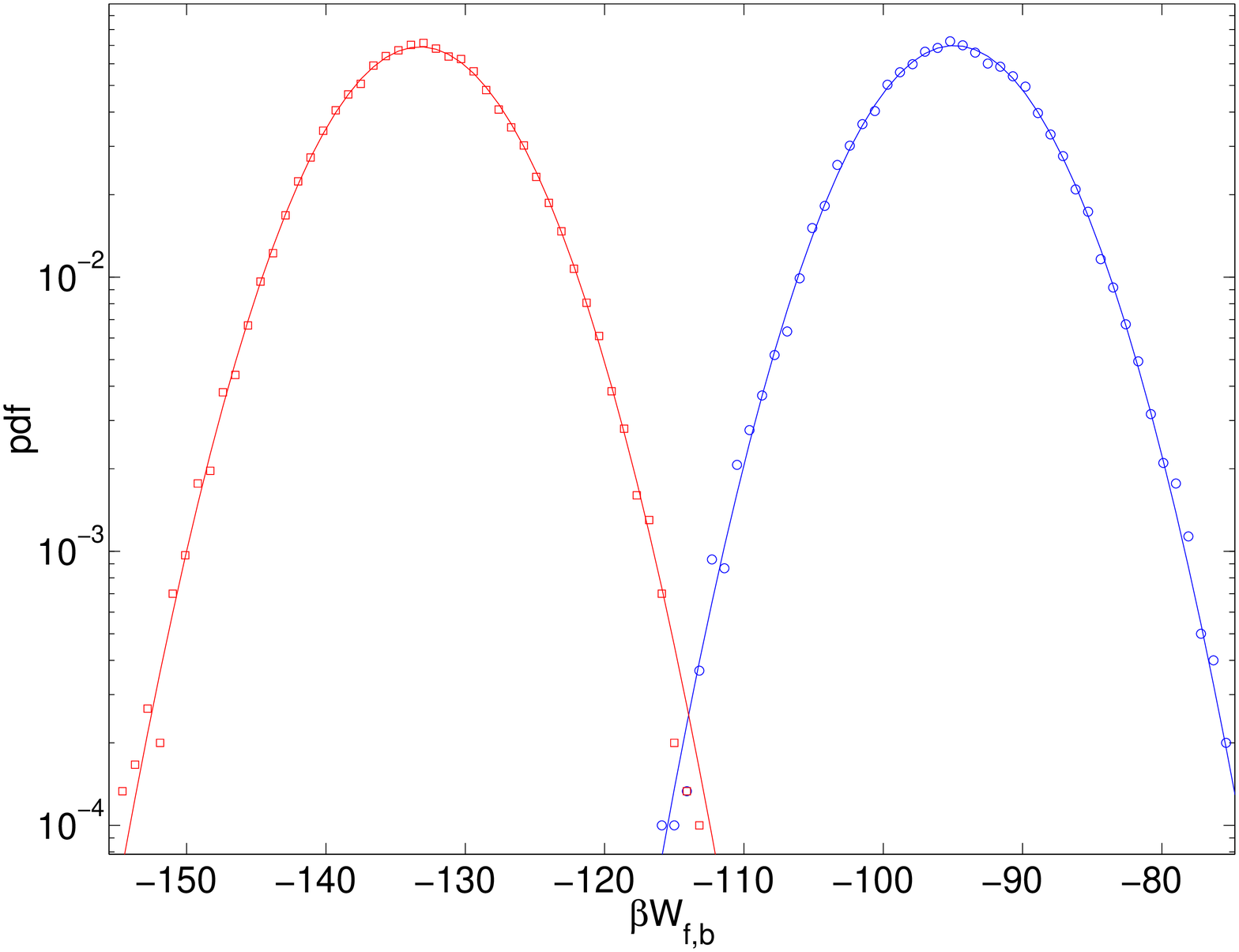}
    \includegraphics[width=7cm, angle=0]{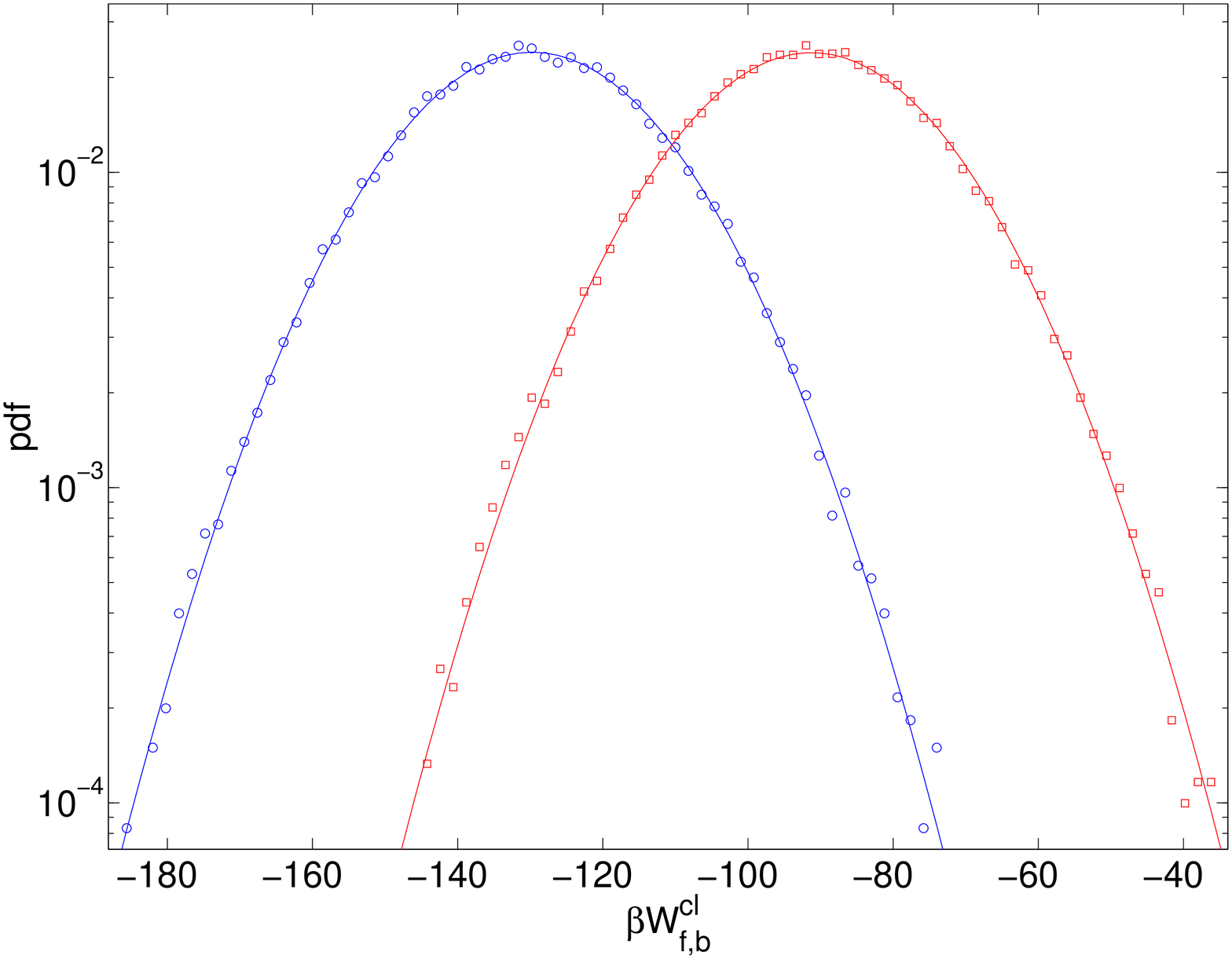}
    \caption{(i) Time evolution of $\theta$ for the case g) in Table \ref{results},
    (ii) Its psd $\sqrt{\langle { \vert \hat{\theta} \vert }^2
    \rangle}$, (iii) $\mathrm{P}_{\mathrm{f}}(W)$ and $\mathrm{P}_{\mathrm{b}}(-W)$,
    (iv) $\mathrm{P}_{\mathrm{f}}(W^{\mathrm{cl}})$ and $\mathrm{P}_{\mathrm{b}}(-W^{\mathrm{cl}})$
    (experimental forward and backward pdfs are represented by $\circ$ and $\Box$ respectively,
    whereas the continuous lines are Gaussian fits)}
    \label{highdrive}
    \end{center}
\end{figure}

Now we drive the oscillator out of equilibrium between two states
$A$ (where $M = 0$) and $B$ (where $M = M_{m} = \mathrm{const}
\neq 0$). The path $\ga$ may be changed by modifying the time
evolution of $M$ between $A$ and $B$. We have chosen either linear
ramps with different rising times $\tau$, see
Fig.\,\ref{driver_fluct_spec_displacement_displacement_pdf_power_class_power}(i),
or half-sinusoids with half-period $\tau$. In the specific case of
our harmonic oscillator, as the temperature is the same in states
$A$ and $B$, the free energy difference of the oscillator alone is
$\Delta F_0 = \Delta U = \Big[ \frac{1}{2} C \theta^2 \Big]_A^B =
\Big[ \frac{M^2}{2C} \Big]_A^B$, whereas $\Delta F = \Delta F_0 -
\Big[ \frac{M^2}{C} \Big]_A^B$, i.e. for an harmonic potential
$\Delta F = -\Delta F_0$. Let us first consider the situation
where the cell is filled with oil. The oscillator's relaxation
time is given by the inverse of the line width of the equilibrium
fluctuation spectrum, see Fig.\,\ref{setup_and_fdt_check}(ii),
that is $\tau_{\mathrm{relax}} = 23.5 \textrm{ ms}$. We apply a
torque which is a sequence of linear increasing\,/decreasing ramps
and plateaux, as represented in
Fig.\,\ref{driver_fluct_spec_displacement_displacement_pdf_power_class_power}(i).
We chose different values of the amplitude of the torque $M$
[$22.1$ $11.9$, $6.1$, $4.2$ and $1.2$ pN\,m] and of the rising
time $\tau$ [$199.5$, $20.2$, $65.6$, $99.6$, $2.5$ ms], as
indicated in Table \ref{results} [cases a)\dots g)] (for the case
f) and g) also $C$ has been changed). Thus we can probe either the
reversible (or quasi-static) paths ($\tau \gg
\tau_{\mathrm{relax}}$) or the irreversible ones ($\tau \ll
\tau_{\mathrm{relax}}$). We tune the duration of the plateaux
(which is at least $4\, \tau_{\mathrm{relax}}$) so that the system
always reaches equilibrium in the middle of each of them, which
defines the equilibrium states $A$ and $B$. We see in
Fig.\,\ref{driver_fluct_spec_displacement_displacement_pdf_power_class_power}(ii),
where the angular displacement $\theta$ is plotted as a function
of time [case a)], that the response of the oscillator to the
applied torque is comparable to the thermal noise spectrum. The
psd of $\theta$ is shown in
Fig.\,\ref{driver_fluct_spec_displacement_displacement_pdf_power_class_power}(iii).
Comparing this measure with the FDT prediction obtained in
Fig.\,\ref{setup_and_fdt_check}(ii), one observes that the driver
does not affect the thermal noise spectrum which remains equal to
the equilibrium one. Moreover we plot in
Fig.\,\ref{driver_fluct_spec_displacement_displacement_pdf_power_class_power}(iv)
the pdf of the driven displacement $\theta$ shown on
Fig.\,\ref{driver_fluct_spec_displacement_displacement_pdf_power_class_power}(ii),
which is, roughly speaking, the superposition of two Gaussian
pdfs. From the measure of $M$ and $\theta$, the power injected
into the system $\dot{W}$ can be computed from the definition
given in Eq.\,(\ref{work}), that in this case is $\dot{W} =
-\dot{M} \theta$. Its time evolution, shown in
Fig.\,\ref{driver_fluct_spec_displacement_displacement_pdf_power_class_power}(v),
is quite different from that of the classical power
$\dot{W}^{\mathrm{cl}} = -M \dot{\theta}$, whose time evolution is
plotted in
Fig.\,\ref{driver_fluct_spec_displacement_displacement_pdf_power_class_power}(vi):
$\dot{W}$ is non-zero only for $\dot M \neq 0$ and vice-versa
$\dot{W}^{\mathrm{cl}} \neq 0$ only for $M \neq 0$. From the  time
series of $\dot{W}$ we can compute from Eq.\,(\ref{work}) the
forward and the backward works, $W_\mathrm{f}$ and $W_\mathrm{b}$,
corresponding to the paths $A \to B$ and $B \to A$, respectively.
We also do the same for the classical work. We then compute their
respective pdfs $\mathrm{P}_{\mathrm{f}}(W)$ and
$\mathrm{P}_{\mathrm{b}}(-W)$. These are plotted on
Figs.\,\ref{pdfs}(i,iv) where the bullets are the experimental
data and the continuous lines their fitted Gaussian pdfs. In
Fig.\,\ref{pdfs}, the pdfs of $W$ and $W^{\mathrm{cl}}$ cross in
the case a) at $\beta W \simeq -23.5$, and in the case c) $\beta W
\simeq -6.1$. These values correspond to $\Delta F =
-\frac{M_{m}^2}{2C} = -\Delta F_0$. We find that this result is
true independently of the ratio $\tau / \tau_{\mathrm{relax}}$ and
of the maximum amplitude of $|M|$, $M_{m}$. This has been checked
at the largest $M_{m}$ and the shortest rising time $\tau$ allowed
by our apparatus. Indeed torques with amplitudes larger than  $25
\textrm{ pN\,m}$ and rising time shorter than $2 \textrm{ ms}$
introduce mechanical noises which can be higher than thermal
fluctuations and the check of the JE becomes impossible (see
discussion in the next section). The measurement at very large
$M_{m}$ and very short $\tau$ is shown in Fig.\,\ref{highdrive}.
Also in this case we see that the shape of the thermal noise
spectrum is not perturbed by the driving. The pdfs remain Gaussian
but the distance between ${\langle W \rangle}_{\mathrm{f}}$ and
${\langle W \rangle}_{\mathrm{b}}$ is larger and the relative
variance much smaller than at low amplitude. The crossing point of
the pdfs of $W$ occurs at a value where the statistics is very
poor. However the two Gaussian fits crosses at $\Delta F$. The
pdfs of $W^{\mathrm{cl}}$ also crosses at the right values.
The experimental results are summarized in Table \ref{results},
where the computed $\Delta U = {M_{m}^2 \over 2C}$ is in good
agreement with the values obtained by the crossing points of the
forward and backward pdfs, that is $\Delta F_{\times} + \Phi$ for
$\mathrm{P}(W)$ and $-\Delta W_{\times}^{\mathrm{cl}}$ for
$\mathrm{P}(W^{\mathrm{cl}})$. Finally inserting the values of
$W_{\mathrm{f}}$ and $W_{\mathrm{b}}$ in Eq.\,(\ref{JE}) we
directly compute $\Delta F_{\mathrm{f}}$ and $\Delta
F_{\mathrm{b}}$ from the JE. As it can be seen in Table
\ref{results}, the values of $\Delta F_0$ obtained from the JE,
that is either $-(\Delta F_{\mathrm{f}} + \Phi)$ or $-(\Delta
F_{\mathrm{b}} + \Phi)$, agree (see sec.3.3) with the computed
$\Delta U$ within experimental errors, that are about 5\% on
$\Delta U$ (see sec.3.3). The JE works well either when $\tau \gg
\tau_{\mathrm{relax}}$ or in the critical case f) and g) where
$\tau \ll \tau_{\mathrm{relax}}$. The other case we have studied
is a very pathological one. Specifically, the oscillator is in
vacuum and has a resonant frequency $f_0 = 353 \textrm{ Hz}$ and a
relaxation time $\tau_{\mathrm{relax}} = 666.7 \textrm{ ms}$. We
applied a sinusoidal torque whose amplitude is either $5.9 \times
10^{-12}$ or $9.4 \times10^{-12} \textrm{ N\,m [cases h) and i) in
Table \ref{results}, respectively]}$. Half a period of the
sinusoid is $\tau = 49.5 \textrm{ ms}$, much smaller than the
relaxation time, so that we never let the system equilibrate.
However, we define the states $A$ and $B$ as the maxima and minima
of the driver. Surprisingly, despite of the pathological
definition of the equilibrium states $A$ and $B$, the pdfs are
Gaussian and the JE is satisfied as indicated in Table
\ref{results}. Moreover, this happens independently of $M_{m}$ and
of the critical value of the ratio $\tau / \tau_{\mathrm{relax}}
\ll 1$. Finally, we indicated in Table \ref{results} the value
$\Delta F_{\circlearrowleft}$ which is the free energy computed
from the JE if one considers the ``loop process'' from $A$ to $A$
(the same can be done from $B$ to $B$ and the results are
quantitatively the same). In principle this value should be zero,
but in fact it is not since we have about 3\% error in the
calibration of the torque $M$ and on $C$.
\begin{table}[!t]
    \begin{center}
    \begin{tabular}{|c|c|c|c|c|c|c|c|}
    \hline
    $\tau / \tau_{\mathrm{relax}}$ & $M_{m}$ & $-\beta[\Delta F_{\mathrm{f}} + \Phi]$ & $\beta[\Delta F_{\mathrm{b}} + \Phi]$ & $-\beta[\Delta F_{\times} + \Phi]$ & $-\beta W_{\times}^{\mathrm{cl}}$ & $\beta \Delta U$ & $ |\beta \Delta F_{\circlearrowleft}|$ \\
    \hline\hline
    $8.5^{\textrm{ a)}}$ & $11.9$ & $23.5$ & $23.1$ & $23.5$ & $23.4$ & $23.8$ & $1.0$\\
    $0.85^{\textrm{ b)}}$ & $6.1$ & $6.6$ & $6.1$ & $6.0$ & $6.6$ & $6.1$ & $1.0$\\
    $3.5^{\textrm{ c)}}$ & $6.1$ & $6.1$ & $5.9$ & $6.5$ & $6.1$ & $6.1$ & $0.4$\\
    $2.8^{\textrm{ d)}}$ & $4.2$ & $2.8$ & $2.6$ & $3.2$ & $2.9$ & $2.7$ & $0.3$\\
    $4.2^{\textrm{ e)}}$ & $1.2$ & $0.21$ & $0.20$ & $ 0.22 $ & $ 0.21 $ & $0.22$ & $0.04$\\
    \hline \hline
    $0.11^{\textrm{ f)}}$ & $11.8$ & $33$ & $30.8$ & $32.54$ & $31.15$ & $31.4$ & $3.6$\\
    $0.11^{\textrm{ g)}}$ & $22.1$ & $117.6$ & $110.5$ & $114$ & $110.1$ & $111$ & $15.1$\\
    \hline \hline
    $0.07^{\textrm{ h)}}$ & $5.9$ & $10.3$ & $10.0$ & $10.1$ & $10.1$ & $10.3$ & $0.4$\\
    $0.07^{\textrm{ i)}}$ & $9.4$ & $67.4$ & $65.5$ & $66.8$ & $66.4$ & $67.5$ & $2.4$\\
    \hline
    \end{tabular}
    \vspace{0.5cm}
    \caption{Free energies of cases a)\dots g) defined in the text (the values of $M_{m}$ are in pN\,m).
    $\Delta U = {M_{m}^2 \over 2C}$ is the computed expected
    value. Notice that $C = 7.5 \times 10^{-4} \mathrm{\ N\,m\,rad^{-1}}$ for cases a)-e),
    $C = 5.5 \times 10^{-4} \mathrm{\ N\,m\,rad^{-1}}$
    for cases f)-g) and $C = 1.6 \times 10^{-4} \mathrm{\ N\,m\,rad^{-1}}$ for cases h)-i)}
    \label{results}
    \end{center}
\end{table}
\section{Jarzynski equality and the Langevin equation}
\begin{figure}
    \begin{center}
{\hspace{3cm} (i) \hspace{4cm} (ii)  \hspace{3.5cm} (iii)} \\
    \includegraphics[width=4.8cm, angle=0]{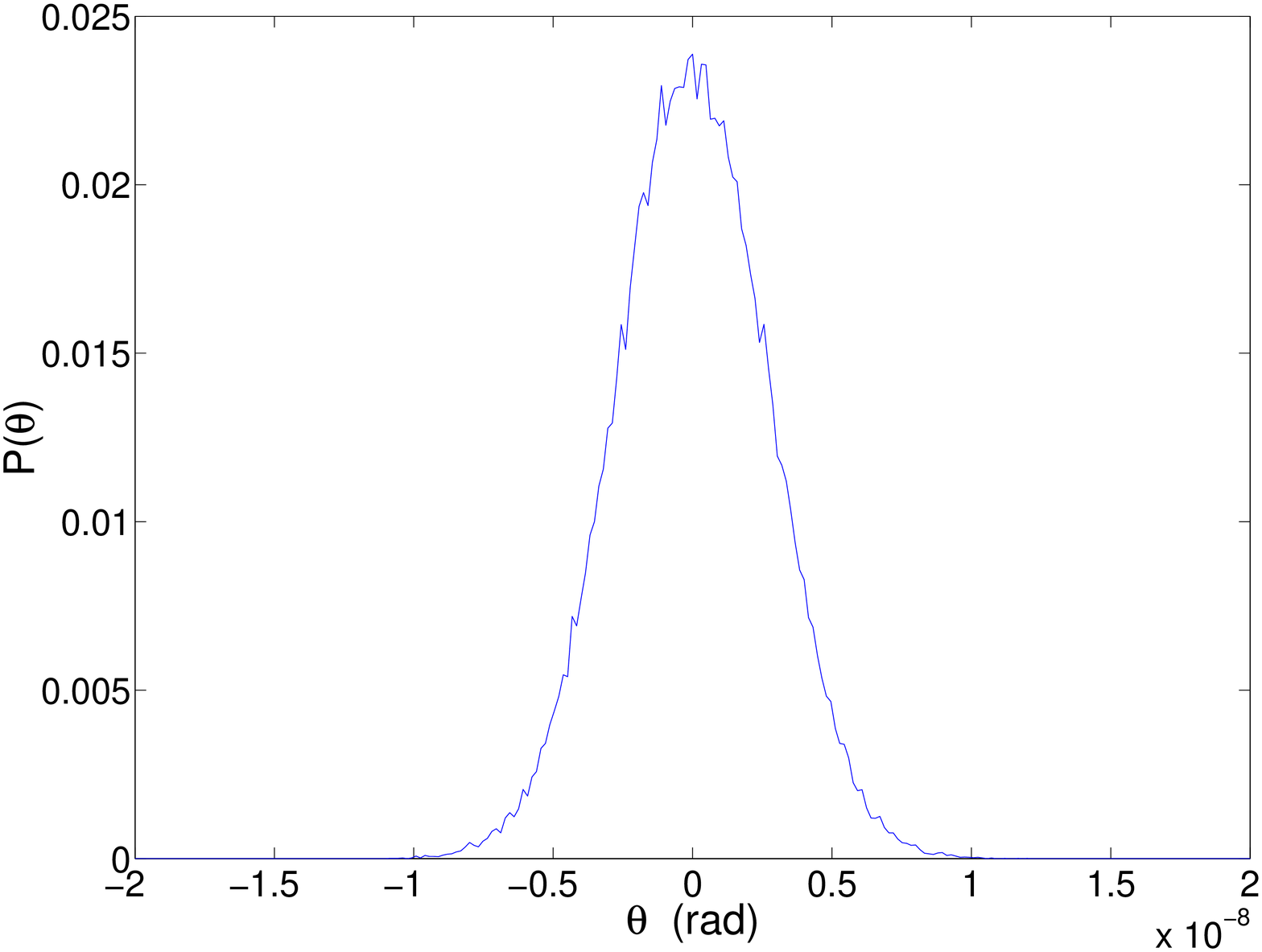}
    \includegraphics[width=4.8 cm, angle=0]{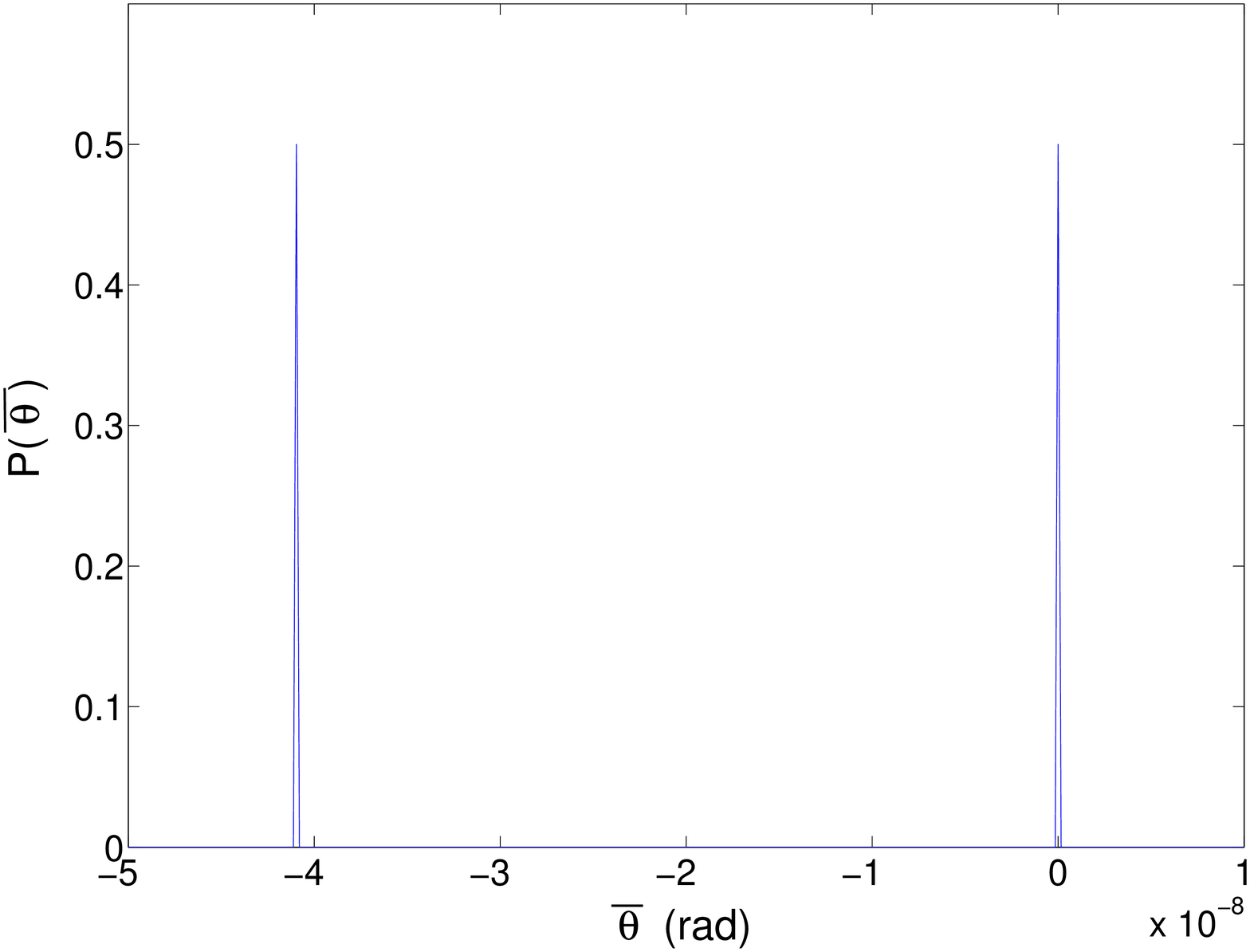}
    \includegraphics[width=4.8cm, angle=0]{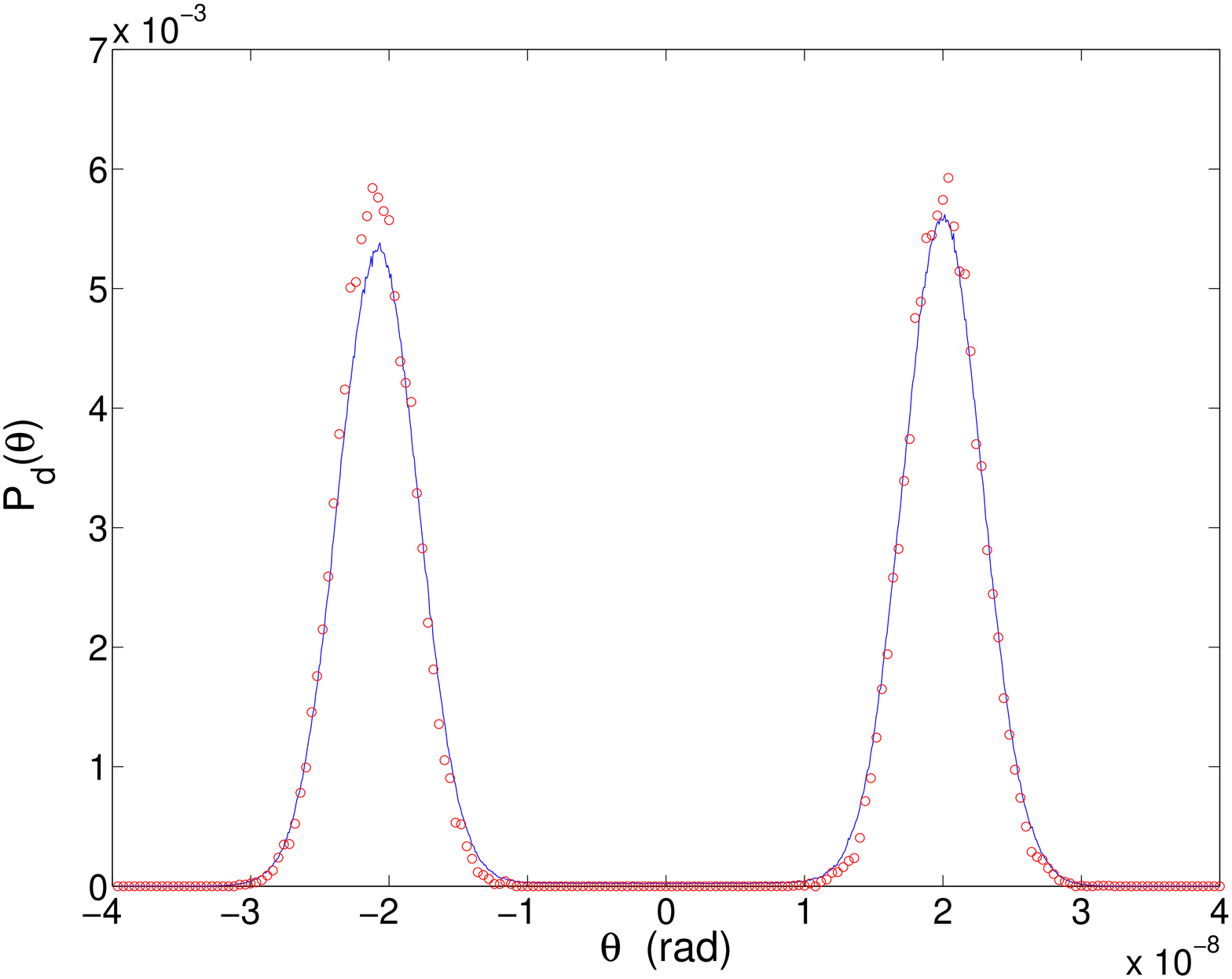}
    \caption{(i) PDF  of $\theta$ when the driving is off.
    (ii) PDF $P(\overline{\theta})$  of the mean response of $\theta$
     to the torque, measured for the case g) of Table
 \ref{results}.
 (iii) PDF $P_d(\theta)$ of $\theta$ measured for the case g) of Table
 \ref{results} ($\circ$) and computed by the convolutions of the
 PDF plotted in (i) and (ii) (continuous line).}
    \label{convolution}
    \end{center}
\end{figure}

In the previous section we have seen that even at very large
driving torque, two properties of the systems remain unchanged,
specifically:  the thermal noise amplitude and statistics are not
modified by the presence of the large forcing, that is FDT is
still valid and the fluctuation pdf remains Gaussian. As we have
already discussed,  the fact that FDT is still valid is easily
understood by comparing the noise spectrum of
fig.\ref{setup_and_fdt_check}-ii) without driving with those with
driving  in
fig.\ref{driver_fluct_spec_displacement_displacement_pdf_power_class_power}-ii)
and fig.\ref{highdrive}ii). We clearly see that the  driving does
not change the shape of the noise spectrum, once the pics
corresponding to the driving  have been subtracted. It is also
easy to show that  the pdf $P_d(\theta)$ of $\theta$, with the
driving, is the convolution product of the pdf $P(\theta)$,
without driving, times
 $P(\overline{\theta})$ which is the mean response of $\theta$ to the
 driving torque.
 As an example we show in fig.\ref{convolution}a) $P(\theta)$
without driving. The PDF
 $P(\overline{\theta})$ of the mean response of $\theta$, measured for the case g) of Table
 \ref{results}, is plotted in
 fig.\ref{convolution}b). This PDF  is close to the sum of
  two delta functions. In fig.\ref{convolution}c) we show, for the case g) of Table
 \ref{results}, the
  directly measured $P_d(\theta)$ (circles) and that computed by
  the convolution of $P(\overline{\theta}))$ with $P(\theta)$.
  The agreement is excellent. This result shows that the statistical
  properties of the thermal noise are those of equilibrium even when the
  system is driven very fast.

\noindent Starting from these  experimental evidences, we can show
that for the Langevin equation the JE is always satisfied
independently of $M_{m}$ and $\tau$. Let us consider the equation
for the harmonic oscillator, Eq.\,(\ref{eqofmotion}), which well
describes our experimental system. In the case of a viscous
damping we rewrite Eq.\,(\ref{eqofmotion}) as
\begin{equation}
    I_{\mathrm{eff}}\,\ddot{\theta} + \nu \,\dot{\theta} + C\,\theta = M + \eta,
    \label{eqoscillator}
\end{equation}
where $\eta$ is the thermal noise amplitude. For $M(t)$ we
consider the kind of waveform used in the experiments:
\begin{eqnarray}
    M(t) & = & {M_{m} \,t \over \tau} \qquad \mathrm{for} \qquad 0 < t < \tau,\\
    & = & M_{m} \qquad \mathrm{for} \qquad t > \tau.
\end{eqnarray}
\subsection{The overdamped case}
In order to compute $\Delta F$ using the JE when the system is
driven from $A$ ($M = 0$) to $B$ ($M = M_{m} \neq 0$), we first
consider the overdamped case when the inertial term is negligible,
that is $I_{\mathrm{eff}}\simeq 0$:
\begin{equation}
    \dot{\theta} + \tau_0^{-1}\,\theta = (C\,\tau_0)^{-1} (M + \eta),
    \label{eqoverdamped}
\end{equation}
where $\tau_0 = \nu / C$. If $\eta$ is a thermal noise, then when
$M = 0$ the spectrum of the thermal fluctuations $\delta \theta$
of $\theta$ can be computed from FDT:
\begin{equation}
    \langle { \vert \hat{\theta} \vert }^2 \rangle
    = \frac{4 k_B T\,\tau_0}{C (1 + \tau_0^2 \, \omega^2)}
    \label{fdt1}
\end{equation}
As a consequence the autocorrelation function of $ \theta$ on a
time interval $\delta \tau $ is
\begin{equation}
    R_\theta(\delta \tau) =
    \frac{k_B T }{C}\,\exp{\Bigl(-{|\delta\tau|\over\tau_0}\Bigr)}.
    \label{corr1}
\end{equation}
The pdf of $\theta$ is  Gaussian. The work to drive the system
from $A$ to $B$ computed using Eq.\,(\ref{work}) becomes in this
case
\begin{equation}
    W = - {M_{m} \over \tau} \int_{0}^{\tau} \theta \d t
    \label{W1}
\end{equation}
Thus to compute $W$ we need only the solution of Eq.\,(\ref{eqoverdamped})
for $0 < t < \tau$. If we neglect the noise, then the mean solution is
\begin{equation}
    \overline{\theta} = {M_{m} \over \tau\,C} \Bigl[t + \tau_0\,
    \exp{\Bigl(-{t \over \tau_0}\Bigr)} -\tau_0 \Bigr] \qquad \mathrm{for} \qquad 0 < t < \tau.
    \label{solution1}
\end{equation}
We now consider that in the experiment the statistical properties
of the thermal fluctuations $\delta\theta$ are not modified by the
driving. Therefore $\theta$ can be decomposed in the sum of an
average part $\overline{\theta}$ given by Eq.\,(\ref{solution1})
plus the fluctuating part $\delta \theta$, that is
\begin{equation}
    \theta = \overline{\theta} + \delta\theta.
    \label{somtheta}
\end{equation}
As a consequence the work can also be decomposed in a similar manner
\begin{equation}
    W = \overline{W} + \delta W = -{M_{m} \over \tau} \Biggl[
    \int_{0}^{\tau} \overline{\theta} \d t + \int_{0}^{\tau} \delta \theta \d t
    \Biggr].
    \label{somW}
\end{equation}
As the integral of a Gaussian variable is still Gaussian then the
fluctuations of $W$ remain Gaussian too. As a consequence, to
compute $\Delta F$ we can use Eq.\,(\ref{eqgausF}), where $\langle W \rangle$
is straightforward computed using Eqs.\,(\ref{solution1}) and (\ref{somW})
\begin{equation}
    \langle W \rangle = \overline{W} = -{M_{m} \over \tau}\,
    \int_{0}^{\tau}\overline{\theta} \d t =
    -{1 \over C}{\left({M_{m} \over \tau}\right)}^2
    \Bigl[{(\tau - \tau_0)^2 \over 2} - \tau_0^2
    \exp{\Bigl(-{\tau \over \tau_0}\Bigr)} + {{\tau_0}^2 \over 2} \Bigr].
    \label{meanW}
\end{equation}
We now have to compute $\sigma_W^2 = \langle {(\delta W})^2
\rangle = ({M_{m} / \tau})^2 \, \langle y^2(\tau) \rangle$ where
\begin{equation}
    y(\tau) = \int_0^{\tau} \delta\theta \d t.
\end{equation}
The variance of $y$ can be computed \cite{Papoulis} taking into account that
\begin{equation}
    \langle y^2(\tau) \rangle = \int_0^{\tau} \int_0^{\tau} R_\theta(t_1 - t_2) \d t_1 \d t_2.
    \label{variancey}
\end{equation}
Using this equation and Eq.\,(\ref{corr1}) we get
\begin{equation}
    \sigma_W^2 = \left({M_{m} \over \tau}\right)^2 \langle y^2(\tau) \rangle =
    {2 k_B T \tau_0 \over C} \left({M_{m} \over \tau}\right)^2 \Bigl[
    \tau - \tau_0 + \tau_0 \exp{\Bigl(-{\tau \over \tau_0} \Bigr)} \Bigr].
    \label{varianceW}
\end{equation}
Taking into account that fluctuations of $W$ are Gaussian, we replace
the results of Eqs.\,(\ref{meanW}) and (\ref{varianceW}) in
Eq.\,(\ref{eqgausF}), and finally we get
\begin{equation}
    \Delta F = \langle W \rangle - {\sigma_W^2 \over 2 k_B T} =
    -{M_{m}^2 \over 2 C},
\label{DeltaF1}
\end{equation}
that is the expected value. It is important to notice that this
equation gives the exact result independently of the rising time
of the external applied torque. This is important because it
shows that the in cases where the conditions a) and b) are
verified, the JE gives the right result independently of the path to go
from $A$ to $B$, which can be a very irreversible one.
\subsection{The harmonic oscillator}
We may now repeat the calculation for the harmonic oscillator of
Eq.\,(\ref{eqoscillator}). In such a case let us introduce the
following notations
\begin{equation}
    \alpha = {\nu \over 2 I_{\mathrm{eff}}},\qquad
    \alpha^2 + \psi^2 = {C \over I_{\mathrm{eff}}},
    \label{notation1}
\end{equation}
and
\begin{equation}
    \cos{\varphi} = {\alpha \over \sqrt{\alpha^2 + \psi^2}},\qquad
    \sin{\varphi} = {\psi \over \sqrt{\alpha^2 + \psi^2}}.
    \label{notation2}
\end{equation}
With this notation the mean solution of Eq.\,(\ref{eqoscillator}) in absence of
noise, with initial conditions $\theta (0) = \dot{\theta} (0) = 0$, is
\begin{equation}
    \bar{\theta} = {M_{m} \over \tau \psi} \Bigl[
    \exp{(-\alpha t)}\, \sin{(\psi t + 2 \varphi)} + \psi\,t -
    \sin{2\varphi} \Bigr] \qquad \mathrm{for} \qquad 0 < t < \tau.
    \label{solution2}
\end{equation}
The correlation function of the thermal fluctuations $\delta\theta$
become in this case
\begin{equation}
    R_{\theta}(\delta\tau)= {k_B T \over C \sin{\varphi}}\,
    \exp{(-\alpha |\delta \tau|)}\, \sin{(\psi|\delta\tau| + \varphi)}.
\label{correlation2}
\end{equation}
We now proceed as in the overdamped case and we compute $\langle W \rangle$
and $\sigma_W^2$. Using Eq.\,(\ref{solution2}) and Eq.\,(\ref{somW}) we get
\begin{equation}
    \langle W \rangle = \left( {M_{m} \over \tau \psi} \right)^2
    {\sin{\varphi} \over C } \left[ \exp{(-\alpha \tau)}\,
    \sin{(\psi \tau + 3\varphi)} - \sin{3\varphi} - {(\psi \tau)^2 \over 2
    \sin{\varphi}} + 2 \psi \tau \cos{\varphi} \right]
    \label{meanW2}
\end{equation}
To compute the variance of $W$, we insert Eq.\,(\ref{correlation2}) in
Eq.\,(\ref{variancey}) and we obtain
\begin{eqnarray}
    \sigma_W^2 & = & \left({M_{m} \over \tau}\right)^2 \langle y^2(\tau) \rangle
    \notag\\
    & = & \left({M_{m} \over \tau}\right)^2 {2 k_B T \sin{\varphi} \over C \psi^2}
    \left[ \exp{(-\alpha \tau)}\,\sin{(\psi \tau + 3 \varphi)} - \sin{3\varphi}
    + 2 \psi \tau \cos{\varphi} \right].
    \label{varianceW2}
\end{eqnarray}
As the fluctuations of $W$ are Gaussian, we insert Eqs.\,(\ref{meanW2}) and
(\ref{varianceW2}) in Eq.\,(\ref{eqgausF}), and we obtain
\begin{eqnarray}
    \Delta F = \langle W \rangle - {\sigma_W^2 \over 2 k_B T} = -{M_{m}^2 \over 2 C},
    \label{DeltaF2}
\end{eqnarray}
which is the expected results. Notice that in this case too the
result is independent on the path. Thus the JE gives the right result
of $\Delta F$ for the Langevin equation with an harmonic potential.

\section{Discussion and Conclusions}

\subsection{Experimental limits of JE and CR}
Before concluding we want to discuss  the limits on the use of JE
and CR in an experiment  where the dynamics can be either exactly
described or well approximated by a Langevin equation. From the
values of $\langle W \rangle$ and of $\sigma_W$, computed in sect.
4 for the overdamped case and for the harmonic oscillator case, we
see that if $\tau$ is kept constant then $\langle W \rangle
\propto M_{m}^2$ whereas $\sigma_W \propto M_{m}$. Furthermore as
$2\,\Delta F = {\langle W \rangle}_{\mathrm{f}} - {\langle W
\rangle}_{\mathrm{b}}$ is proportional to $M_{m}^2$, this means
that the distance between the maxima of
$\mathrm{P}_{\mathrm{f}}(W)$ and $\mathrm{P}_{\mathrm{b}}(-W)$
increases with $M_{m}^2$ and the relative width $|\sigma_W /
\langle W \rangle|$ of $\mathrm{P}(W)$ decreases as $1/M_{m}$. As
a consequence, the probability of finding experimental values of
$W$ close to the crossing points also decreases as $1/M_{m}$. This
effect has been seen on the experimental $\mathrm{P}(W)$ plotted
in Fig.\,\ref{highdrive}. From a practical point of view, this
means that when $\Delta F \gg k_B T$ the pdfs will never cross,
for reasonable values of the work pdf. As a consequence the
experimental test of JE and CR in systems with a strongly
irreversible driving force, as proposed in ref.\cite{cohen}, will
be very difficult to test. Indeed at such a large and fast
driving,  the JE and CR may fail, either for deep theoretical
reasons or simply  because the statistics is too poor. At the
moment we are not able to answer to this question. Furthermore,
for real (macroscopic) systems it is reasonable to think that when
the external noise becomes much larger than the thermal noise, the
JE cannot be used. However, if the pdfs of $W^{\mathrm{cl}}$
remain Gaussian, then the crossing point
$W_{\times}^{\mathrm{cl}}$ gives the right result.

\subsection{Summary and discussion of the results}

In conclusion  we have used a driven torsion pendulum to test
experimentally  the accuracy of  JE and CR. By varying the
amplitude and the rising time of the driving torque of about one
order of magnitude, we clearly demonstrate the validity and the
robustness of the JE and CR in an isothermal process, at least
when the work fluctuations are Gaussian and when the harmonic
approximation is relevant for the system. We have checked the
generality of the results on a driven Langevin equation which well
describes the dynamics of the oscillator. Using the experimental
observations that the equilibrium properties of the thermal noise
are not modified by the driving force and that the force
fluctuations are Gaussian, we have shown that the JE gives the
right result independently of $M_{m}$ and $\tau$. Unfortunately
these results do not fully alight the theoretical debate, because
our conditions are close to those pointed out in
Ref.\,\cite{cohen} for the validity of the JE. Recently, Ritort
and coworkers have used the JE and CR  to estimate $\Delta F$ in
an experiment of RNA stretching where the oscillator's coupling is
non-linear and the work fluctuations are non-Gaussian
\cite{ritort_told_us}. It would be interesting to check these
results on a more simple and controlled system. We are currently
working on the experimental realization of such a non-linear
coupling, for which $\Delta F \neq -\Delta F_0$.

We have also shown both analytically and experimentally that
$|\sigma_W / \langle W \rangle|$ decreases as $1/M_{m}$. This
observation makes the practical use of the JE and CR rather
unrealistic for very large $M_{m}$ as the statistics needed to get
a reliable result will be very large. This means, as we have
already discussed at the end of the previous section, that the JE
cannot be applied for macroscopic systems when the external noise
becomes much larger than the thermal noise.

Going back to the estimate of $\Delta F$ we have seen that the
more accurate and reliable $\Delta F$ estimator is given by the
crossing points $\Delta F_{\times}$ and
$W_{\times}^{\mathrm{cl}}$, because they are less sensitive to the
extreme fluctuations which may perturb the convergence of the JE.
Starting from this observation, we propose a new  method to
compute $\Delta F$ in the case of Gaussian fluctuations of
$W^{\mathrm{cl}}$. Indeed, for Gaussian pdfs
$W_{\times}^{\mathrm{cl}}$ remains an excellent estimator even in
cases where the JE and the CR could not hold, for example when the
environmental noise cannot be neglected.

Finally we want to stress that our results, although limited to
the Gaussian case, show that it is possible to measure tiny work
fluctuations in a macroscopic system. As a consequence it opens a
lot of perspective to use the JE, the CR and the recent theorems
on dissipated work (see for example \cite{cvz}) to characterize
the slow relaxation towards equilibrium in more complex systems,
for example aging materials such as glasses or
gels \cite{crisanti}.\\

The authors thank L. Bellon, E.G.D. Cohen, N. Garnier, C.
Jarzynski, F. Ritort and L. Rondoni for useful discussions, and
acknowledge P. Metz, M. Moulin, F. Vittoz, C. Lemonias and P.-E. Roche
 for technical support. This work has been partially
supported by the {\sc{Dyglagemem}} contract of EEC.
\end{document}